\documentclass[12pt]{article}
\usepackage{epsfig}
\usepackage{feynmf}
\usepackage{a4p}
\usepackage{cite}

\begin{document}
\renewcommand{\thefootnote}{\arabic{footnote}}
\renewcommand{\figurename}{\protect\it Figure}
\renewcommand{\tablename}{\protect\it Table}
\def\ssee{s_{\rm ee}}
\def\ssgg{s_{\gamma\gamma}}
\def\sqee{\sqrt{s}_{\rm ee}}
\def\sqeeb{\sqrt{s}_{\protect\bf\rm ee}}
\def\sqgg{\sqrt{s}_{\gamma\gamma}}
\def\gg{\gamma\gamma}
\def\Wvis{W_{\rm vis}}
\def\Wecal{W_{\rm ECAL}}
\def\qmax{Q^2_{\rm max}}
\def\qmin{Q^2_{\rm min}}
\def\ee{\mbox{e}^+\mbox{e}^-}
\def\fg{f_{\gamma/{\rm e}}}
\def\sgg{\sigma_{\gamma\gamma}}
\def\see{\sigma_{\mbox{\footnotesize e}^+\mbox{\footnotesize e}^-}}
\def\zav{\langle z_0 \rangle}
\def\Qav{Q}
\def\EECAL{E_{\rm ECAL}}
\def\EFD{E_{\rm FD}}
\def\ESIW{E_{\rm SiW}}
\def\WECAL{W_{\rm ECAL}}
\def\nch{n_{\rm ch}}
\newcommand{\sleq} {\raisebox{-.6ex}{${\textstyle\stackrel{<}{\sim}}$}}
\newcommand{\sgeq} {\raisebox{-.6ex}{${\textstyle\stackrel{>}{\sim}}$}}
\def\ETJET{E^{\rm jet}_{\rm T}}
\def\pt{p_{\rm T}}
\def\ETBAR{\bar{E}^{\rm jet}_{\rm T}}
\def\EJET{E^{\rm jet}}
\def\PZJET{p^{\rm jet}_z}
\def\ETi{E_{{\rm T}_i}}
\def\ETMIN{E^{\rm min}_{\rm T}}
\def\thetamax{\Theta_{\rm max}}
\def\qqbar{\mbox{q}\overline{\mbox{q}}}
\def\qbar{\overline{q}}
\def\xg{x_{\gamma}}
\def\xgp{x_{\gamma}^+}
\def\xgm{x_{\gamma}^-}
\def\xgpm{x_{\gamma}^{\pm}}
\def\etajet{\eta^{\rm jet}}
\def\phijet{\phi^{\rm jet}}
\def\Zzero{\ifmmode {{\mathrm Z}^0} \else {${\mathrm Z}^0$} \fi}
\def\ppbar{\overline{\mbox p}\mbox{p}}
\def\pz{\phantom{0}}
\def\pzz{\phantom{00}}
\def\pzzz{\phantom{.00}}
\def\shat{M_{\rm jj}}
\def\PTMIA{p_{\rm t}^{\rm mi}}
\def\cost{\cos\theta^{*}}
\hyphenation{PHOJET}
\hyphenation{PYTHIA}

\begin{titlepage}
\begin{center}{\large   EUROPEAN ORGANISATION FOR NUCLEAR RESEARCH
}\end{center}\bigskip
\begin{flushright}
       CERN-EP/98-113   \\ 16th July 1998
\end{flushright}
\bigskip\bigskip\bigskip\bigskip
\begin{center}
\boldmath\huge\bf   
Di-Jet Production in 
Photon-Photon Collisions at 
$\sqeeb=161$ and $172$~GeV\unboldmath
\end{center}\bigskip\bigskip
\bigskip\bigskip
\begin{center}{\LARGE The OPAL Collaboration
}\end{center}\bigskip\bigskip\bigskip
\bigskip\begin{center}{\large  Abstract}\end{center}
Di-jet production is studied in collisions of quasi-real photons
radiated by the LEP beams at e$^+$e$^-$ centre-of-mass energies
$\sqee=161$ and $172$~GeV. The jets are reconstructed
using a cone jet finding algorithm.
The angular distributions of direct and double-resolved
processes are measured and compared to the predictions of leading order
and next-to-leading order perturbative QCD.
The jet energy profiles are also studied.
The inclusive two-jet cross-section is measured as a function of
$\ETJET$ and $|\etajet|$ and
compared to next-to-leading order perturbative QCD calculations. 
The inclusive two-jet cross-section as a function of $|\etajet|$
is compared to the prediction of the leading order Monte Carlo
generators PYTHIA and PHOJET. The Monte Carlo predictions are
calculated with different parametrisations of the
parton distributions of the photon.
The influence of the `underlying event'
has been studied to reduce the model dependence of the
predicted jet cross-sections from the Monte Carlo generators.

\bigskip
\bigskip

%
%
%

\bigskip\bigskip
\begin{center}
{\bf \large
}\end{center}

\begin{center}{\large
(To be submitted to European Physics Journal C)
}\end{center}
\end{titlepage}

\begin{center}{\Large        The OPAL Collaboration
}\end{center}\bigskip
\begin{center}{
G.\thinspace Abbiendi$^{  2}$,
K.\thinspace Ackerstaff$^{  8}$,
G.\thinspace Alexander$^{ 23}$,
J.\thinspace Allison$^{ 16}$,
N.\thinspace Altekamp$^{  5}$,
K.J.\thinspace Anderson$^{  9}$,
S.\thinspace Anderson$^{ 12}$,
S.\thinspace Arcelli$^{ 17}$,
S.\thinspace Asai$^{ 24}$,
S.F.\thinspace Ashby$^{  1}$
% ...
}\end{center}\bigskip
\bigskip
$^{  1}$School of Physics and Astronomy, University of Birmingham,
Birmingham B15 2TT, UK
\newline
$^{  2}$Dipartimento di Fisica dell' Universit\`a di Bologna and INFN,
I-40126 Bologna, Italy
\newline
$^{  3}$Physikalisches Institut, Universit\"at Bonn,
D-53115 Bonn, Germany
\newline
$^{  4}$Department of Physics, University of California,
Riverside CA 92521, USA
\newline
$^{  5}$Cavendish Laboratory, Cambridge CB3 0HE, UK
\newline
$^{  6}$Ottawa-Carleton Institute for Physics,
Department of Physics, Carleton University,
Ottawa, Ontario K1S 5B6, Canada
\newline
$^{  7}$Centre for Research in Particle Physics,
Carleton University, Ottawa, Ontario K1S 5B6, Canada
\newline
$^{  8}$CERN, European Organisation for Particle Physics,
CH-1211 Geneva 23, Switzerland
\newline
$^{  9}$Enrico Fermi Institute and Department of Physics,
University of Chicago, Chicago IL 60637, USA
\newline
$^{ 10}$Fakult\"at f\"ur Physik, Albert-Ludwigs-Universit\"at,
D-79104 Freiburg, Germany
\newline
$^{ 11}$Physikalisches Institut, Universit\"at
Heidelberg, D-69120 Heidelberg, Germany
\newline
$^{ 12}$Indiana University, Department of Physics,
Swain Hall West 117, Bloomington IN 47405, USA
\newline
$^{ 13}$Queen Mary and Westfield College, University of London,
London E1 4NS, UK
\newline
$^{ 14}$Technische Hochschule Aachen, III Physikalisches Institut,
Sommerfeldstrasse 26-28, D-52056 Aachen, Germany
\newline
$^{ 15}$University College London, London WC1E 6BT, UK
\newline
$^{ 16}$Department of Physics, Schuster Laboratory, The University,
Manchester M13 9PL, UK
\newline
$^{ 17}$Department of Physics, University of Maryland,
College Park, MD 20742, USA
\newline
$^{ 18}$Laboratoire de Physique Nucl\'eaire, Universit\'e de Montr\'eal,
Montr\'eal, Quebec H3C 3J7, Canada
\newline
$^{ 19}$University of Oregon, Department of Physics, Eugene
OR 97403, USA
\newline
$^{ 20}$CLRC Rutherford Appleton Laboratory, Chilton,
Didcot, Oxfordshire OX11 0QX, UK
\newline
$^{ 22}$Department of Physics, Technion-Israel Institute of
Technology, Haifa 32000, Israel
\newline
$^{ 23}$Department of Physics and Astronomy, Tel Aviv University,
Tel Aviv 69978, Israel
\newline
$^{ 24}$International Centre for Elementary Particle Physics and
Department of Physics, University of Tokyo, Tokyo 113, and
Kobe University, Kobe 657, Japan
\newline
$^{ 25}$Institute of Physical and Environmental Sciences,
Brunel University, Uxbridge, Middlesex UB8 3PH, UK
\newline
$^{ 26}$Particle Physics Department, Weizmann Institute of Science,
Rehovot 76100, Israel
\newline
$^{ 27}$Universit\"at Hamburg/DESY, II Institut f\"ur Experimental
Physik, Notkestrasse 85, D-22607 Hamburg, Germany
\newline
$^{ 28}$University of Victoria, Department of Physics, P O Box 3055,
Victoria BC V8W 3P6, Canada
\newline
$^{ 29}$University of British Columbia, Department of Physics,
Vancouver BC V6T 1Z1, Canada
\newline
$^{ 30}$University of Alberta,  Department of Physics,
Edmonton AB T6G 2J1, Canada
\newline
$^{ 31}$Duke University, Dept of Physics,
Durham, NC 27708-0305, USA
\newline
$^{ 32}$Research Institute for Particle and Nuclear Physics,
H-1525 Budapest, P O  Box 49, Hungary
\newline
$^{ 33}$Institute of Nuclear Research,
H-4001 Debrecen, P O  Box 51, Hungary
\newline
$^{ 34}$Ludwigs-Maximilians-Universit\"at M\"unchen,
Sektion Physik, Am Coulombwall 1, D-85748 Garching, Germany
\newline
\bigskip\newline
$^{  a}$ and at TRIUMF, Vancouver, Canada V6T 2A3
\newline
$^{  b}$ and Royal Society University Research Fellow
\newline
$^{  c}$ and Institute of Nuclear Research, Debrecen, Hungary
\newline
$^{  d}$ and Department of Experimental Physics, Lajos Kossuth
University, Debrecen, Hungary
\newline
$^{  e}$ on leave of absence from the University of Freiburg
\newline

\section{Introduction}
We present a study of di-jet production in photon-photon collisions at
$\sqee=161$ and $172$~GeV with an integrated luminosity of
20 pb$^{-1}$. The cone jet finding algorithm was used to reconstruct
jets.
The production of di-jet events in the collision of two quasi-real photons
can be used to study the structure of the photon and to test
QCD predictions. At $\ee$ colliders the photons are emitted
by the beam electrons\footnote{Positrons
are also referred to as electrons}. Most of the photons carry only
a small squared four-momentum, $Q^2$, and can be considered to be
quasi-real ($Q^2 \approx 0$).
Accordingly, the electrons are scattered with very small angles and
are not detected. Events where one or both scattered electrons are
detected are vetoed (``anti-tagged'').

The interactions of the photons can be modelled by assuming that
each photon can either interact directly or appear resolved
through its fluctuations into hadronic components.
The interaction of two photons can be classified either as a direct
process where two bare photons interact, a single-resolved process
where a bare photon interacts with a parton (quark or gluon) 
of the other photon or a double-resolved process where partons of
both photons interact.
The possibility to distinguish between direct and resolved processes
in di-jet events has already been demonstrated by OPAL 
at $\sqee=130-136$~GeV~\cite{bib-opalgg}.
Depending on the type of photon-photon interaction different matrix
elements for the QCD scattering process contribute.
These matrix elements have been
calculated in leading order (LO)~\cite{bib-maelm} and
next-to-leading order (NLO)~\cite{bib-klasen2}.
The matrix element of the scattering between two bare photons is
the one for the process $\gg \rightarrow \qqbar$. In double-resolved
processes the matrix
elements of quark-quark, gluon-quark and gluon-gluon scattering
are involved~\cite{bib-maelm}. These calculations predict
different distributions of the parton scattering angle
in the centre-of-mass system of the colliding particles.
In hadron-hadron and photon-hadron interactions similar QCD predictions
have already been confirmed~\cite{bib-cdf,bib-zeusgp}.

The investigation of the internal structure
of jets gives insight into the transition between a parton produced in
a hard process and the observable hadrons which originate from the
fragmentation process~\cite{bib-sdellis}. 
The dependence of the jet shapes
on QCD parton radiation calculated in the Leading-Log 
Approximation (LLA), and the differences
between the jet shapes of
direct and resolved processes, have recently been measured
at HERA~\cite{bib-zeusshape}.

The measurement of inclusive jet cross-sections in $\gg$ and
$\gamma$p interactions
can constrain the gluonic content of the photon~\cite{bib-amy,bib-hera}.
This is done by comparing
the jet cross-sections to the LO QCD models PYTHIA and PHOJET, 
using different parametrisations
of the parton distribution functions of the photon, and to NLO
QCD calculations~\cite{bib-kleinwort,bib-aurenche}.
PYTHIA and PHOJET also model the so-called `underlying event'
by multiple interactions involving partons from the remnants of the same two
initial photons, whereas the NLO QCD calculations
do not take into account such effects.
In the models, the contribution from multiple interactions
to the jet cross-sections has to be tuned to increase
sensitivity to the parton distributions of the photon.
In contrast to deep inelastic electron-photon scattering~\cite{bib-OPALF2}, 
which in leading order is only sensitive to
the quark content of the photon, the gluon content of the photon
can be tested directly in the resolved 
interaction of two almost real photons, where the fraction of
gluon initiated processes is large.

Inclusive one-jet and two-jet cross-sections in photon-photon collisions
have previously been
measured at an $\ee$ centre-of-mass energy of $\sqee=$ 58 GeV at
TRISTAN \cite{bib-amy,bib-topaz} and at an $\ee$ centre-of-mass
energy of $\sqee=130$ and 136 GeV at LEP~\cite{bib-opalgg}.
This paper extends our analysis at lower energies in which a similar strategy
was used. Jets are studied in a wider kinematic range and with
higher integrated luminosity. In addition, we present new results
on angular distributions, jet shapes and energy flows.

\section{Process kinematics}
\label{sec-kine}
The properties of the interacting photons are described by their
negative squared four-mo\-mentum transfers, $Q_{i}^2$.
Each $Q_i^2$ is related to the electron
scattering angle $\Theta'_i$ relative to the beam direction by
\begin{equation}
Q_i^2=-(p_i-p'_i)^2\approx 2E_i E'_i(1-\cos\Theta'_i),
\label{eq-q2}
\end{equation} 
where $p_i$ and $p'_i$ are the four-momenta of the beam
electrons and the scattered electrons, respectively,
and $E_i$ and $E'_i$ are their energies.
Events with detected scattered electrons (single-tagged or
double-tagged events) are excluded from
the analysis. This anti-tagging condition defines an upper
limit on $Q_{i}^2$ for both photons. 
This condition is met when the
scattering angle $\Theta'$ of the electrons is
less than $\thetamax'$, where $\thetamax'$ is the angle
between the beam-axis and the inner edge of the detector.
The squared invariant mass of the hadronic final state
\begin{equation}
W^2=\left(\sum_{h}E_h\right)^2-\left(\sum_{h}\vec{p}_h\right)^2
\end{equation}
is calculated by summing over the energies, $E_{h}$, and 
momenta, $\vec{p}_h$, 
of all final state hadrons.
The spectrum of photons with an energy fraction $y$ of the electron
beam may be obtained
by the Equivalent Photon Approximation (EPA) \cite{bib-wwa}:
$$\fg(y)=
\frac{\alpha}{2\pi}\left(\frac{1+(1-y)^2}{y}
\log\frac{\qmax}{\qmin}
-2m^{2}_{\rm e} y\left( \frac{1}{\qmin}-\frac{1}{\qmax} \right)\right),$$
with $\alpha$ being the electromagnetic coupling constant.
The minimum kinematically allowed squared four-momentum 
transfer, $\qmin$, is determined by 
the electron mass $m_{\rm e}$: 
$$\qmin=\frac{m_{\rm e}^2y^2}{1-y}.$$
The effective maximum four-momentum transfer
$\qmax$ is given by the anti-tagging
condition, i.e.~the requirement that both electrons remain undetected.

\section{The OPAL detector}
A detailed description of the OPAL detector
can be found in Ref.~\cite{opaltechnicalpaper}, and
therefore only a brief account of the main features relevant
to the present analysis will be given here.
 
The central tracking system is located inside 
a solenoidal magnet which
provides a uniform axial magnetic field of 0.435~T along the beam
axis\footnote{In the OPAL coordinate system 
  the $z$ axis points in the direction of the e$^-$ beam. The
  polar angle $\theta$, the azimuthal angle $\phi$
  and the radius $r$ denote the usual spherical coordinates.}.
The magnet is surrounded in the barrel region ($|\cos\theta|<0.82$)
by a lead glass electromagnetic
calorimeter (ECAL) and a hadronic sampling calorimeter (HCAL).  
Outside the HCAL, the detector is surrounded by muon
chambers. There are similar layers of detectors in the 
endcaps ($0.82<|\cos\theta|<0.98$).
The small angle region from 47 to 140 mrad
around the beam pipe on both sides
of the interaction point is covered by the forward calorimeters (FD)
and the region from 25 to 59 mrad by the silicon tungsten luminometers 
(SW)~\cite{bib-siw}.
From 1996 onwards, relevant to the data presented in this paper,
the lower boundary of the acceptance has been  increased to 33 mrad
following the installation of a low angle shield to protect the
central tracking system against possible synchrotron radiation.

Starting with the innermost components, the
tracking system consists of a high precision silicon
microvertex detector, a vertex
drift chamber, a large volume jet chamber with 159 layers of axial
anode wires and a set of $z$ chambers measuring the track coordinates
along the beam direction. 
The transverse momenta, $\pt$, of tracks are measured with a precision 
parametrised by
$\sigma_{\pt}/\pt=\sqrt{0.02^2+(0.0015\cdot \pt)^2}$ ($\pt$ in GeV/$c$)
in the central region. In this paper ``transverse''
is always defined with respect to the $z$ axis.
The jet chamber also provides 
measurements of the energy loss, ${ \rm d} E/ {\rm d}x$, 
which are used for particle identification~\cite{opaltechnicalpaper}.

The barrel and endcap sections of the ECAL  are
both constructed from lead glass blocks with a depth of
$24.6$ radiation lengths in the barrel region and more than 
$22$ radiation lengths in the endcaps. 
The FD consist of cylindrical lead-scintillator calorimeters with a depth of   
24 radiation lengths divided azimuthally into 16 segments.  
The electromagnetic energy resolution is about
$18\%/\sqrt{E}$, where $E$ is in GeV.                                  
The SW detectors consist
of 19 layers of silicon detectors and 18
layers of tungsten, corresponding to a total of 22 radiation
lengths. Each silicon layer consists of 16 wedge
shaped silicon detectors. The electromagnetic energy resolution is about
$25\%/\sqrt{E}$ ($E$ in GeV).  

\section{Event selection and jet finding}
\label{sec-evsel}

Two-photon events are selected with the following set of cuts:
\begin{itemize}
\item
The sum of all energy deposits
in the ECAL and the HCAL has to be less than 45 GeV.
Calorimeter clusters have to pass an energy threshold of 100 MeV in the
barrel section
or 250 MeV in the endcap section for the ECAL and 
of 600 MeV for the barrel and endcap section of the HCAL.
\item
The visible invariant mass measured
in the ECAL has to be greater than 3 GeV.
\item 
The missing transverse energy of the event
measured in the ECAL and the FD has to be less than 5 GeV.
For a FD cluster to be counted its
energy has to be larger than 1 GeV.
\item 
At least 5 tracks must have 
been found in the tracking chambers.
A track is required to have a minimum transverse momentum
of 120 MeV/$c$, at least 20 hits in the central jet chamber,
and the innermost hit of the track 
must be within a radius of 60 cm with respect to the $z$ axis.
The distance of the point of closest approach to the origin
in the $r\phi$ plane must be
less than 30 cm in the $z$ direction and less than
2 cm in the $r\phi$~plane. 
Tracks with a momentum error larger than the momentum itself
are rejected if they have fewer than 80 hits. 
The number of measured
hits in the jet chamber must be more than half of the number of possible hits,
where the number of possible hits
is calculated from the polar angle $\theta$ of the track, assuming 
that the track has no curvature. 
\item
To remove events with scattered electrons in the FD or
in the SW calorimeters,
the total energy sum measured in the FD has to
be less than 50 GeV and the total energy sum measured in the
SW calorimeter has to be less than 35 GeV.
A cluster in the SW calorimeter is accepted if it has
an energy of more than 1 GeV.
These cuts also reduce the contamination from multihadronic 
annihilation events with
their thrust axis close to the beam direction.
\item
In order to estimate the $z$ position of the primary vertex, we calculate
the error-weighted average  $\zav$ of the $z$ coordinates of all 
tracks at the point of closest approach to the origin in the
$r\phi$ plane.
The background due to beam-gas interactions is
reduced by requiring $|\zav|<10$~cm and $|\Qav| \le 3$, where 
$\Qav$ is the net charge of an event calculated from adding the
charges of all tracks.
\item
To remove beam-wall events the radial distance of the primary vertex
from the beam axis has to be less than 3 cm.
\end{itemize}

In the cone jet finding algorithm the total transverse energy $\ETJET$
of the jet inside the cone is the scalar sum of the transverse energies
of its components~\cite{bib-opalgg,bib-coneopal}.
In all parts of this analysis, a sum over the particles in the
event or in a jet means a sum over tracks satisfying
the above quality cuts, and over all calorimeters clusters,
including the FD and SW calorimeters.
An algorithm is applied to avoid double-counting of particle momenta
in the central tracking system and their energy deposits
in the calorimeters~\cite{bib-opalgg}. 
The transverse energy $\ETi$ of a particle $i$ is defined relative to the
$z$ axis of the detector with $\ETi=E_i\sin\theta_i$.
For a cone jet to be accepted,
the value of $\ETJET$ must be greater than a certain minimum $\ETMIN$.
The results of the cone jet finding algorithm depend on $\ETMIN$ and
the cone size $R=\sqrt{(\Delta\eta)^2+(\Delta\phi)^2}$ with 
pseudorapidity $\eta=-\ln\tan(\theta/2)$ and azimuthal
angle $\phi$.
Here the values were chosen to be $R=1$ and $\ETMIN=2$~GeV.
The jet pseudorapidity in the laboratory frame is required to
be within $|\etajet|<2$. Monte Carlo studies have shown
that jets with $|\etajet|<2$ are well reconstructed
and are normally fully contained in the detector. This leads to a wider
$\etajet$ acceptance than in our previous analysis~\cite{bib-opalgg}.

We use data corresponding to an integrated luminosity of
9.9~pb$^{-1}$ at $\sqee=161$~GeV and 10.0~pb$^{-1}$ at $\sqee=172$~GeV. 
After applying all cuts and requiring at least
two jets with $\ETJET>3$~GeV and $|\etajet|<2$,
2845 events remain, equally divided between the two centre-of-mass
energies.
For the purpose of this analysis, the difference between the data
taken at $\sqee = 161$~GeV and at 172~GeV 
is small and therefore the distributions for both energies
have been added.
About $12.3\%$ of the di-jet events contain 3 or more jets.
In events with more than
two jets, only the two jets with the highest $\ETJET$ values are taken.

\section{Monte Carlo simulation}
\label{sec-mc}
The Monte Carlo generators PYTHIA 5.722 \cite{bib-pythia,bib-schuler} and 
PHOJET 1.05c \cite{bib-phojet} are used, both based on LO QCD calculations.
These generators have been optimised to describe
$\gamma$p and $\ppbar$ interactions.
The probability of finding a parton in the photon is taken from
parametrisations of the parton distribution functions.
The SaS-1D parametrisation~\cite{bib-sas} is used as default
in PYTHIA and the LO GRV parametrisation~\cite{bib-grv}
as default in PHOJET. 
All possible hard interactions of quarks, gluons and photons
are simulated using LO matrix elements for massless quarks. 
The default value of the cutoff on the transverse momentum of the
two outgoing partons of $\hat{p}_{\rm t}^{\rm min}=1.4$~GeV$/c$ is used.
More details can be found in Ref.~\cite{bib-opalgg}.

The incoming photons in double-resolved events can be viewed
as beams of partons.
For small parton transverse momenta the LO parton
scattering cross-section diverges and becomes larger than the
non-diffractive cross-section as measured in $\gamma {\rm p}$ collisions.
If more than one parton scattering process is allowed in one event
the problem of too large parton cross-section can be solved.
These multiple interactions are calculated as LO QCD processes between
partons of the photon remnants.
In PYTHIA a lower cutoff parameter $\PTMIA$ is introduced, which describes the
transverse momentum of the parton and is set by default to 1.4 GeV$/c$.
The fluctuations of the number of hard interactions are calculated from
a Poisson distribution.
The average number $n_{\rm mi}$ of interactions in double-resolved
di-jet events simulated by PYTHIA is 1.3 for the default
setting $\PTMIA=1.4$~GeV$/c$ using SaS-1D.
PYTHIA and PHOJET use multiple interactions as a component to model
the underlying event.

In PHOJET the $Q^2$ suppression of the total $\gg$ cross-section
is parametrised using Generalised Vector Meson Dominance (GVMD).
A model for the change of soft hadron production and diffraction with
increasing photon virtuality $Q^2$ is also included.  
The photon-photon mode of PYTHIA only simulates the interactions
of real photons with $Q^2=0$. The virtuality of the photons defined
by $Q^2$ enters only through the Equivalent Photon Approximation in
the generation of the photon 
energy spectrum, but the electrons are scattered at zero angle.
This model is not expected to be correct for larger values of $Q^2$.
The contribution of di-jet events with $Q^2>1$~GeV$^2$ and
$\Theta'<33$~mrad generated with the electron-photon mode of PYTHIA
is small and therefore neglected.

The fragmentation of the parton final state is handled 
in both generators by the routines of JETSET 7.408 \cite{bib-pythia}.
Initial and final state parton radiation is included 
based on the LLA.

All signal and background Monte Carlo samples apart
from beam-gas and beam-wall events were generated
with full simulation of the OPAL detector~\cite{bib-gopal}.
They are analysed using the same reconstruction algorithms as
are applied to the data.

The median $Q^2$ of the selected PHOJET events is of the order 
$10^{-4}$~GeV$^2$.
The visible hadronic invariant mass, $\Wvis$, measured with 
all detector information is well described within the errors of
the measurements by the Monte Carlo simulations.
A detailed comparison between
$\Wvis$ and the generated $W$ can be found in Ref.~\cite{bib-opalhad}.

After applying the detector simulation and the
selection cuts to these events,
about $83 \%$ of all generated Monte Carlo events with at least
two hadron jets in the range $\ETJET>3$~GeV and $|\etajet|<2$
are selected. 
The trigger efficiency for all selected Monte Carlo events 
with at least two reconstructed jets in the detector 
is close to $100 \%$.
The number of background events is small, about $1.5 \%$ in total. 
About $1.1 \%$ of the events in the data sample are expected to be
$\ee$ annihilation events with hadronic final states and $0.4 \%$
$\ee\rightarrow\ee\tau^+\tau^-$ events.
No significant background from beam-gas or beam-wall events
is observed.

\section{Properties of direct and resolved processes}
\label{sec-xg}
In LO QCD, neglecting multiple parton interactions,
two hard parton jets are produced in $\gg$ interactions. 
In single- or double-resolved interactions, the two hard parton 
jets are expected to be accompanied by one or two remnant jets.

A pair of variables, $\xgp$ and $\xgm$, can be defined \cite{bib-LEP2}
which specify the fraction of the photon's momentum participating in the
hard scattering:
\begin{equation}
\xgp=\frac{\displaystyle{\sum_{\rm jets=1,2}(E+p_z)}}
 {{\displaystyle\sum_{\rm hadrons}(E+p_z)}} \;\;\;\mbox{and}\;\;\;
\xgm=\frac{\displaystyle{\sum_{\rm jets=1,2}(E-p_z)}}
{\displaystyle{\sum_{\rm hadrons}(E-p_z)}},
\label{eq-xgpm}
\end{equation}
where $p_z$ is the momentum component along the $z$ axis of the
detector and $E$ is the energy of the jets or hadrons. These
variables give some separation between direct
and resolved di-jet events \cite{bib-opalgg}. 

Ideally, for direct events without remnant jets, the total energy
of the event is contained in the two jets, i.e.~$\xgp=1$ and $\xgm=1$,
whereas for single-resolved events either $\xgp$ or $\xgm$ and
for double-resolved events both values, $\xgp$ and $\xgm$,
are expected to be smaller than~1.
Samples with large direct and double-resolved contributions
can be separated by requiring both $\xgp$ and $\xgm$ to be
larger than 0.8 (denoted as $\xgpm > 0.8$) or both
values to be smaller than 0.8 (denoted as $\xgpm < 0.8$),
respectively. Details about the separation between
different event classes can be found in Ref.~\cite{bib-opalgg}.
In the PYTHIA Monte Carlo using the SaS-1D parametrisation 
$86 \%$ of all events in
the region $\xgpm > 0.8$ originate from direct interactions and
$81 \%$ of all events in the region $\xgpm < 0.8$ originate from
double-resolved interactions.

\begin{figure}[htbp]
   \begin{center}
      \mbox{
          \epsfxsize=16.7cm
          \epsffile{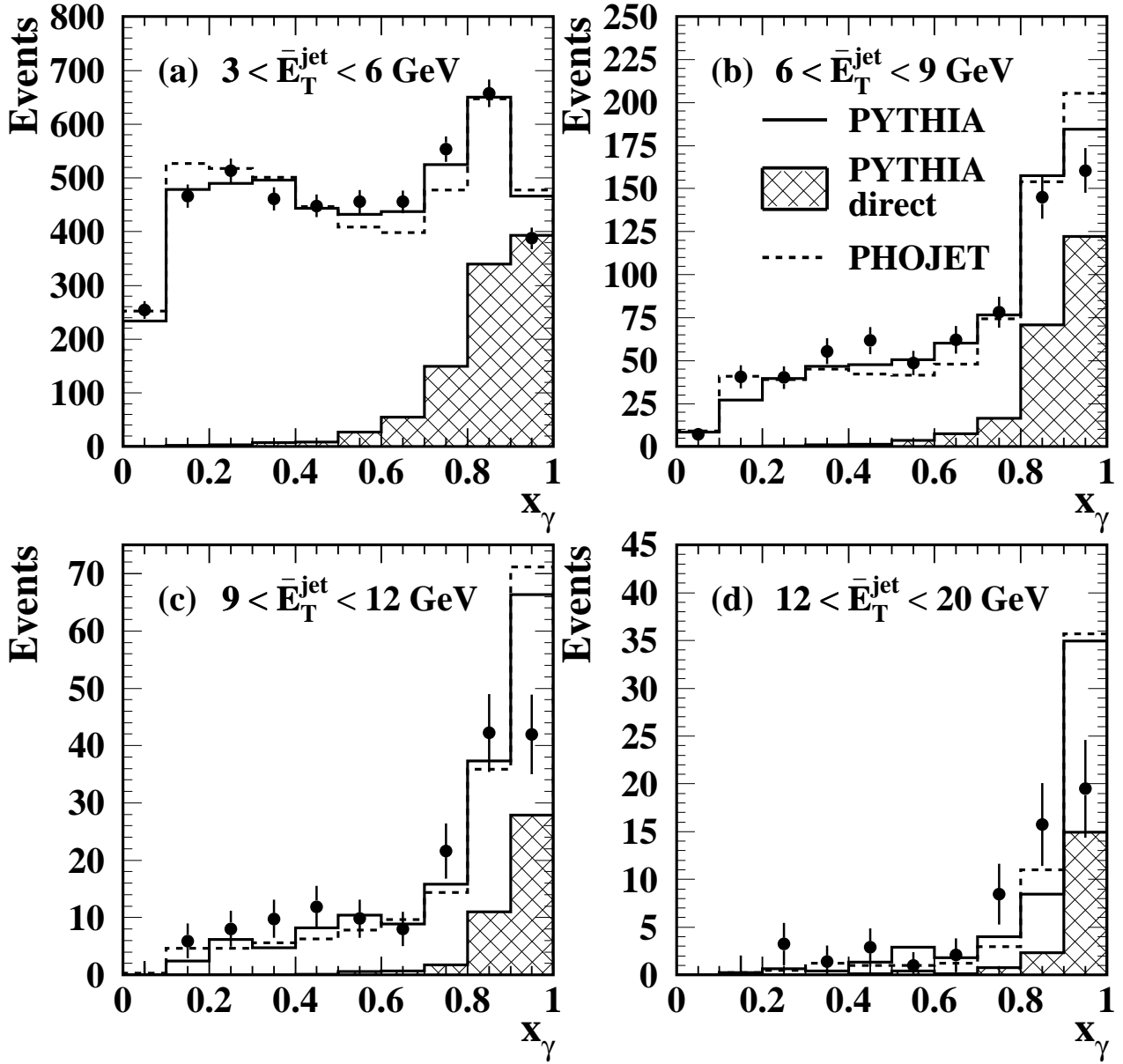}
           }
   \end{center}
\caption{Uncorrected $\xg$ distribution in bins of the mean value of 
$\ETBAR$, where $\ETBAR$ is calculated as the mean value of $\ETJET$
of the two jets with
the highest $\ETJET$. The background has been subtracted using the Monte
Carlo.
The data points are compared to the predictions of
PYTHIA (continuous line) and PHOJET (dashed line). The hatched histogram is
the direct contribution to the PYTHIA events. 
The Monte Carlo histograms are normalised to the number of data events.
Statistical errors only are shown.}
\label{fig-xg}
\end{figure}

The $\xg$ distribution is shown in Figure~\ref{fig-xg} in bins of $\ETBAR$,
where
$$\ETBAR = \frac{E_{\rm T}^{\rm jet1}+E_{\rm T}^{\rm jet2}}{2},$$
is the mean value of the 
transverse energies $E_{\rm T}^{\rm jet1}$ and $E_{\rm T}^{\rm jet2}$ 
of the two jets.
Each event is added
to the plot twice, at the values of $\xgp$ and of $\xgm$.
No correction for selection cuts and detector effects has been applied,
but the background has been subtracted using the Monte Carlo.
The Monte Carlo predictions of PYTHIA and PHOJET are normalised
to the number of events observed in the data. 
The contribution from direct processes, as predicted from PYTHIA,
is also shown. The events from direct processes are concentrated
at high $\xg$ values.
In Fig.~\ref{fig-xg}a, at low $\ETBAR$, the direct part contributes
to about $17 \%$ to the total number of events.
As $\ETBAR$ increases, the $\xg$ distribution
shifts to higher values and the fraction of direct events in the PYTHIA
sample increases to $68\%$ for $12<\ETBAR<20$~GeV (Fig.~\ref{fig-xg}d).
The number of events is underestimated by PYTHIA and PHOJET
by about $25-30 \%$, if the predicted Monte Carlo cross-sections are
taken into account, mainly for $\xg<0.9$.

\section{Angular distributions in direct and resolved events}
\label{sec-angdis}
\begin{figure}[htbp]
   \begin{center}
      \mbox{
          \epsfxsize=16.7cm
          \epsffile{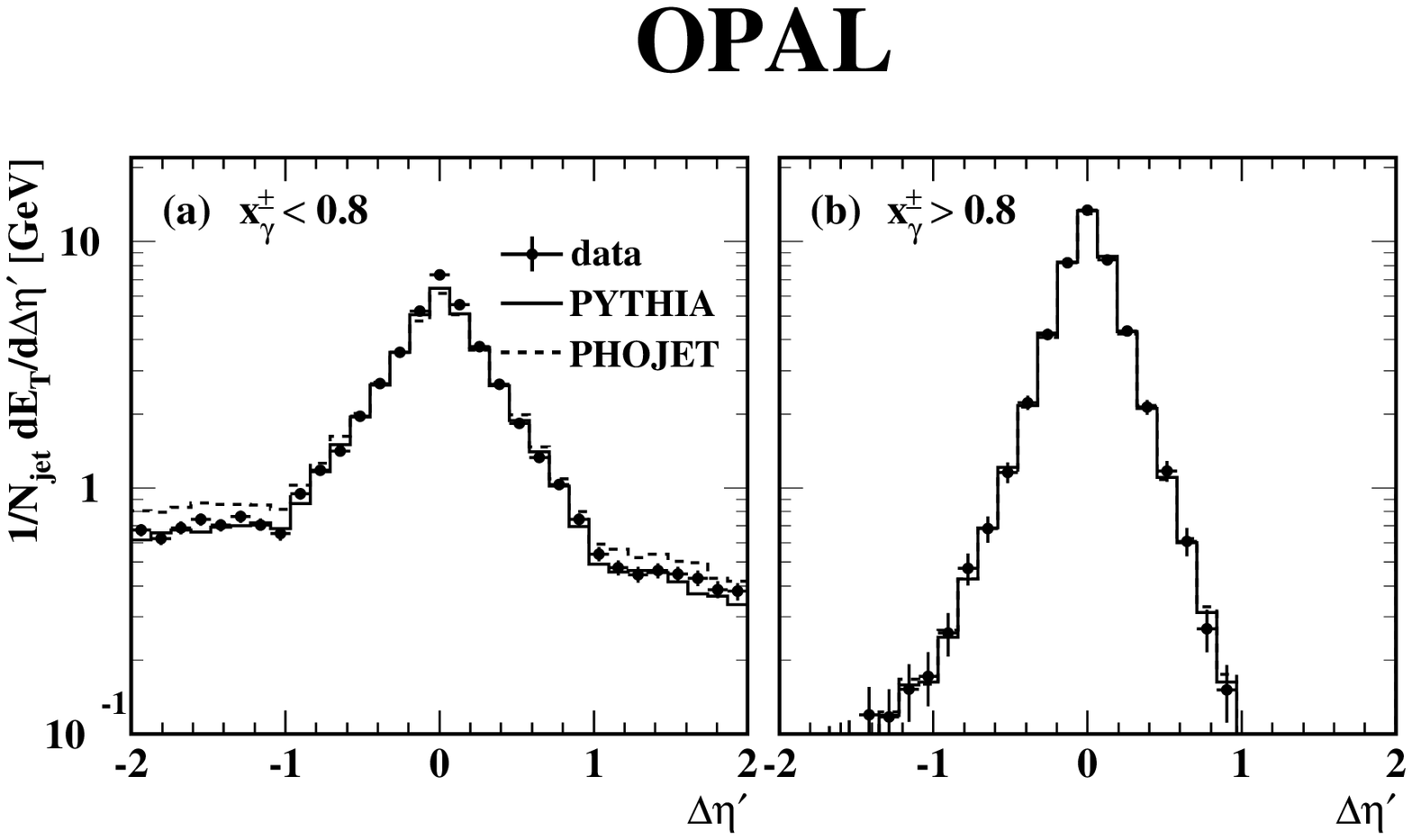}
           }
   \end{center}
\caption{Uncorrected energy flow transverse to the beam direction
measured relative to the direction of each jet in two-jet events
and normalized to the total number of jets, $N_{\rm jet}$, in the sample.
Jets from data samples with a large contribution of (a) double-resolved
and (b) direct events according to their $\xgp$ and $\xgm$ values are
shown. The energy flow is integrated over $|\Delta\phi|<\pi/2$.
Statistical errors only are shown.
The data points are compared to the PHOJET
(continuous line) and PYTHIA (dashed line) simulations.}
\label{fig-jetprof}
\end{figure}

Since the jets in double-resolved events do not contain all of
the hadronic activity it is expected that there will be more energy
flow outside the jets in double-resolved events than in direct events.
Fig.~\ref{fig-jetprof} shows the energy flow transverse to
the beam direction as a function of $\Delta\eta'$
measured with respect to the jet direction for data samples
with different $\xgpm$ cuts. The pseudorapidity difference
is defined by:
$$\Delta\eta'=k(\eta-\etajet),$$
where $\eta$ is the pseudorapidity of the cluster or the track.
The factor $k$ is chosen event-by-event to be
$k=+1$ for events with $\xgp>\xgm$ and $k=-1$ for
events with $\xgp<\xgm$.
As a consequence, there is always more of the remnant found
at $\Delta\eta'<0$
and the enhancement due to the additional transverse energy flow 
observed at negative and positive $\Delta\eta'$ is asymmetric. 
No correction for acceptance or resolution effects has been
applied. The energy flow is integrated over $|\Delta\phi|<\pi/2$.
The jets in the data sample with $\xgpm > 0.8$ (Fig.~\ref{fig-jetprof}a) 
are more collimated
and there is almost no activity outside the jet, whereas the
transverse energy flow of two-jet events with $\xgpm < 0.8$ 
(Fig.~\ref{fig-jetprof}b) 
shows considerable activity outside the jets for $|\Delta\eta'|>1$.
The energy flow outside the jets is well modelled by PYTHIA
whereas PHOJET shows a wider distribution.

In the di-jet centre-of-mass frame one expects different angular
distributions for direct and double-resolved events.
An estimator of the angle $\theta^{*}$ between the jet axis and the
axis of the incoming partons or direct photons
in the di-jet centre-of-mass frame can be formed from the jet
pseudorapidities:
$$\cost=\tanh\left(\frac{\eta^{\rm jet1}-\eta^{\rm jet2}}{2}\right).$$
The two jets are assumed to be collinear in $\phi$ with
$E^{\rm jet1}_{\rm T}=E^{\rm jet2}_{\rm T}$.
Since the ordering of the jets in the detector is
arbitrary, only $|\cost|$ can be measured. 
The matrix elements of elastic parton-parton scattering processes
are known to LO~\cite{bib-maelm}.
For a given parton centre-of-mass energy the cross-sections
vary only with the scattering angle $\theta^{*}$.
The LO direct process $\gg \rightarrow \qqbar$ is mediated
by $t$-channel spin-$\frac{1}{2}$ 
quark exchange which leads to an angular dependence
$\propto (1-\cos^{2}\theta^{*})^{-1}$.
In double-resolved processes all matrix elements
involving quarks and gluons have to be taken into account,
with a large contribution from spin-$0$ gluon exchange. 
After adding up all relevant processes, perturbative QCD
predicts an angular dependence of approximately $\propto
(1-\left|\cost\right|)^{-2}$~\cite{bib-maelm}.

In order to measure the $\cost$ distribution, additional cuts
have to be applied. These cuts minimise kinematic biases and
improve the detector resolution on $\cost$.
The invariant mass of the two-jet system is calculated as
$$M_{\rm jj} \approx
\frac{2 E_{\rm T}^{\rm jet}}{\sqrt{1-\left|\cos\theta^{\ast}\right|^2}}.$$
The cut on $\ETJET>3$~GeV restricts the accessible range of values
of $|\cost|$.
For $M_{\rm jj} < 10$~GeV the number of events 
decreases because of the $\ETJET$ cut.
Requiring $M_{\rm jj}$ to be larger than 12 GeV ensures that the distribution
in the range $|\cost|<0.85$ is not biased by the $\ETJET$ cut.
The Lorentz boost of the two-jet system in the $z$ direction is defined by
$$\bar{\eta}=\frac{\eta^{\rm jet1}+\eta^{\rm jet2}}{2},$$
since in the two-jet centre-of-mass system $\bar{\eta}^*=0$.
Events with $|\bar{\eta}| > 1$ were rejected because the detector resolution
on $|\cost|$ deteriorates significantly for events with $|\bar{\eta}| > 1$.
After additionally requiring $M_{\rm jj} > 12$ GeV and $|\bar{\eta}| < 1$
150 data events remain with $\xgpm>0.8$ and 350 data events
with $\xgpm<0.8$.

Table~\ref{tab-ang} shows the measured cross-sections for
data samples with large direct and large double-resolved
contributions according to the separation with $\xgp$ and $\xgm$.
The cross-section in each bin of $|\cost|$ was corrected according to the
efficiency found for that bin in the PYTHIA and PHOJET samples.
The central value is the mean of the result from PYTHIA and PHOJET.
The systematic uncertainty on the jet cross-sections given
in this section is determined by varying the energy scale
of the ECAL in the Monte Carlo simulation by $\pm 5 \%$.
The dependence on the Monte Carlo models used is taken into
account by adding the difference between the results obtained
with PYTHIA and PHOJET to the systematic error.
The contributions of these two errors to the total systematic
error are of similar size. The systematic errors
due to the luminosity measurement, the trigger efficiency and
the finite number of Monte Carlo events are small in comparison.

\begin{figure}[htbp]
   \begin{center}
      \mbox{
          \epsfxsize=16.7cm
          \epsffile{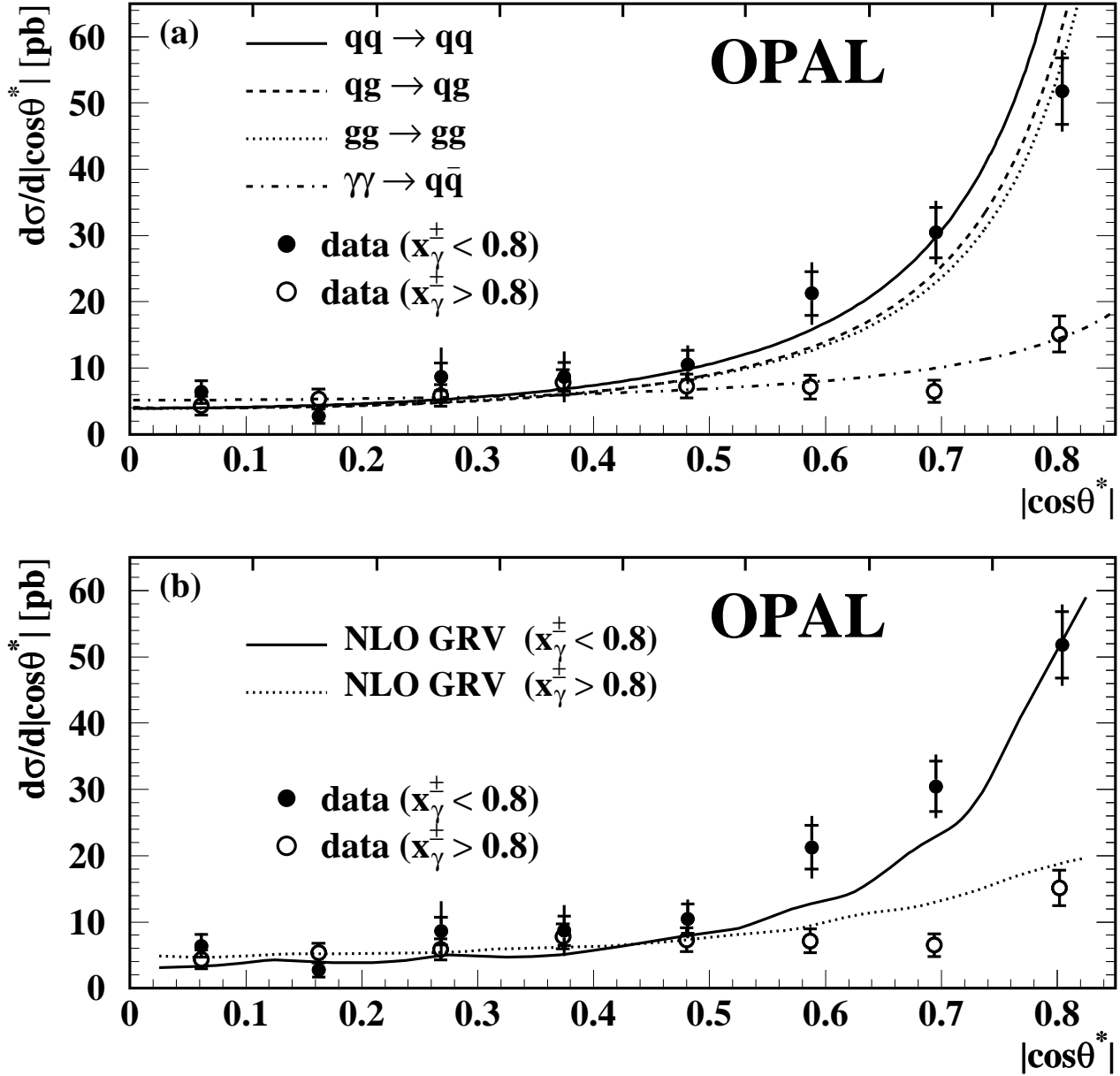}
           }
   \end{center}
\caption{Angular distribution of events with large direct and
large double-resolved contributions according to the separation
with $\xgp$ and $\xgm$. The data are compared (a) to LO QCD matrix element
calculations \cite{bib-maelm} and (b) to NLO QCD calculations
using the GRV parametrisation.
The curves are normalised to the data in the first
three bins.
The open circles show the distribution of events
with $\xgpm>0.8$ and the full circles show the distribution of events
with $\xgpm<0.8$. The inner error bar shows the statistical error
and the outer error bar the statistical and systematic errors added
in quadrature.}
\label{fig-costhst}
\end{figure}

Fig.~\ref{fig-costhst} shows the bin-by-bin
corrected $|\cost|$ distribution
of events with $\xgpm>0.8$ and of events with $\xgpm < 0.8$. 
The abscissae of the data points are plotted according
to the method proposed in Ref.~\cite{bib-stick}.
The predictions of the theoretical parton distributions are integrated
to find the
position of the data points. The error of this position is obtained
using the predictions of the different parton processes. It is smaller
than the line width. 
The predicted curves have been normalised to the data
in the first three bins, in order to compare
the shape of the measured cross-sections as a function of $|\cost|$
with the QCD matrix element calculation and the NLO QCD calculation
The  error bars show the statistical and the systematic errors
added in quadrature. The overall error on the normalisation
of the data points is dominated by the statistical error
in the first three bins, which is about $20\%$.
The events with $\xgpm > 0.8$ show a small rise with $|\cost|$,
whereas the events with $\xgpm < 0.8$ show a much stronger rise
with $|\cost|$, as expected from the QCD calculations.
In the Monte Carlo events about $10 \%$ of the processes with
$\xgpm>0.8$ are double-resolved and about $15 \%$ of the processes with
$\xgpm<0.8$ are direct.
The $|\cost|$ distribution is not much affected by these impurities,
since the Monte Carlo double-resolved events with $\xgpm>0.8$ show
a smaller rise with $|\cost|$ than the double-resolved events with
$\xgpm<0.8$ and
the Monte Carlo direct events with $\xgpm<0.8$ show a stronger rise
with $|\cost|$ than the direct events with $\xgpm>0.8$.

In Fig.~\ref{fig-costhst}a the points for $\xgpm<0.8$ lie close to
the predictions of a QCD matrix element calculation of the interaction
of quarks or gluons in the photon~\cite{bib-maelm}.
The matrix elements with a relevant contribution to the
cross-section where anti-quarks are involved instead of quarks
show a similar behaviour to the examples shown.
The points for $\xgpm>0.8$ are comparable with the results of a
calculation of the process $\gg \rightarrow \qqbar$.
The QCD matrix element
calculations agree well with the data samples with
large direct and large double-resolved contribution.

The data points are compared in Figure~\ref{fig-costhst}b
to a NLO perturbative QCD calculation~\cite{bib-klasen2}.
The contribution of the different processes to all
double-resolved events depends on the parametrisation of the
parton distribution functions.
This calculation uses the NLO GRV parametrisation 
and was repeated for the kinematic conditions
of this analysis.
The shape of the data points and the NLO calculation agrees well.
However, before normalisation
the predicted cross-section is a factor of two too high for
the direct events and about $50 \%$ too small for double-resolved events.
A NLO QCD calculation using the
GS photon structure function~\cite{bib-gs} (not shown) 
shows a similar behaviour.

\section{Jet shapes}
\label{sec-shape}
The internal structure of jets produced in photon-photon interactions
is studied at the hadron level. The jet shape is characterised by the 
fraction of a jet's transverse energy ($\ETJET$) that lies inside
an inner cone of radius $r$ concentric with the jet defining
cone \cite{bib-sdellis}:
\begin{equation}
  \psi(r) = \frac{1}{N_{\rm jet}} \sum_{\rm jets} 
\frac{E_{\rm T}(r)}{E_{\rm T}(r=R)},
\end{equation}
where $E_{\rm T}(r)$ 
is the transverse energy within the inner cone of radius $r$
and $N_{\rm jet}$ is the total number of jets in the sample. By definition,
$\psi(r=R)=1$.

To check the relative importance of parton radiation and fragmentation
in the formation of a jet the parton-shower in the LLA
as implemented in PYTHIA~5.7 \cite{bib-pythia} has been used.
The jet shape is affected both by fragmentation and gluon radiation. 

The jet shapes are 
corrected to the hadron level using the
Monte Carlo event samples with single-resolved, double-resolved and
direct processes. The corrected jet shapes are denoted by
$\psi(r)$ and refer to jets at the hadron level with a cone radius
$R=1$. The reconstructed jet shapes
are corrected for acceptance effects and the finite detector resolution.
The correction factors
also take into account the selection criteria and the purity and efficiency
of the jet reconstruction. The corrected jet shapes are determined
bin-by-bin as
$$\psi(r) = C(r) \cdot \psi_{\rm det}(r),$$
where the correction factors $C(r)$ are defined as
$$C(r) = \frac{\psi^{\rm MC}_{\rm had}(r)}{\psi^{\rm MC}_{\rm det}(r)}.$$
Since correction factors are different for single-resolved, double-resolved and
direct processes, the correction factors have been calculated
for each process separately in each $\ETBAR$ interval.
For direct processes $C(r)$ is close to $1$ for all $r$, whereas
for single- and double-resolved processes $C(r)$ lies in the range
from 0.8 to 1.2 for $r=0.2$ and approaches $1$ as $r$ increases.
The relative contributions
of the different processes have been estimated by comparing the
$\xg$ distribution of the data events with PYTHIA (see Figure~\ref{fig-xg})
and changing the contributions to get a correct description.
In the lowest $\ETBAR$ bin $(3<\ETBAR<6$~GeV) PYTHIA describes
the $\xg$ distribution well and the contributions of the different
processes were not modified. For the bins in the range
$6<\ETBAR<20$~GeV, the contribution of double-resolved events was 
increased and the contribution of direct events has been reduced, whereas
the contribution of single-resolved events was left unchanged.
The correction factor $C(r)$ in each $\ETBAR$ bin is calculated
according to the relative contribution of each process determined
in this way.

\begin{figure}[htbp]
   \begin{center}
      \mbox{
          \epsfxsize=16.7cm
          \epsffile{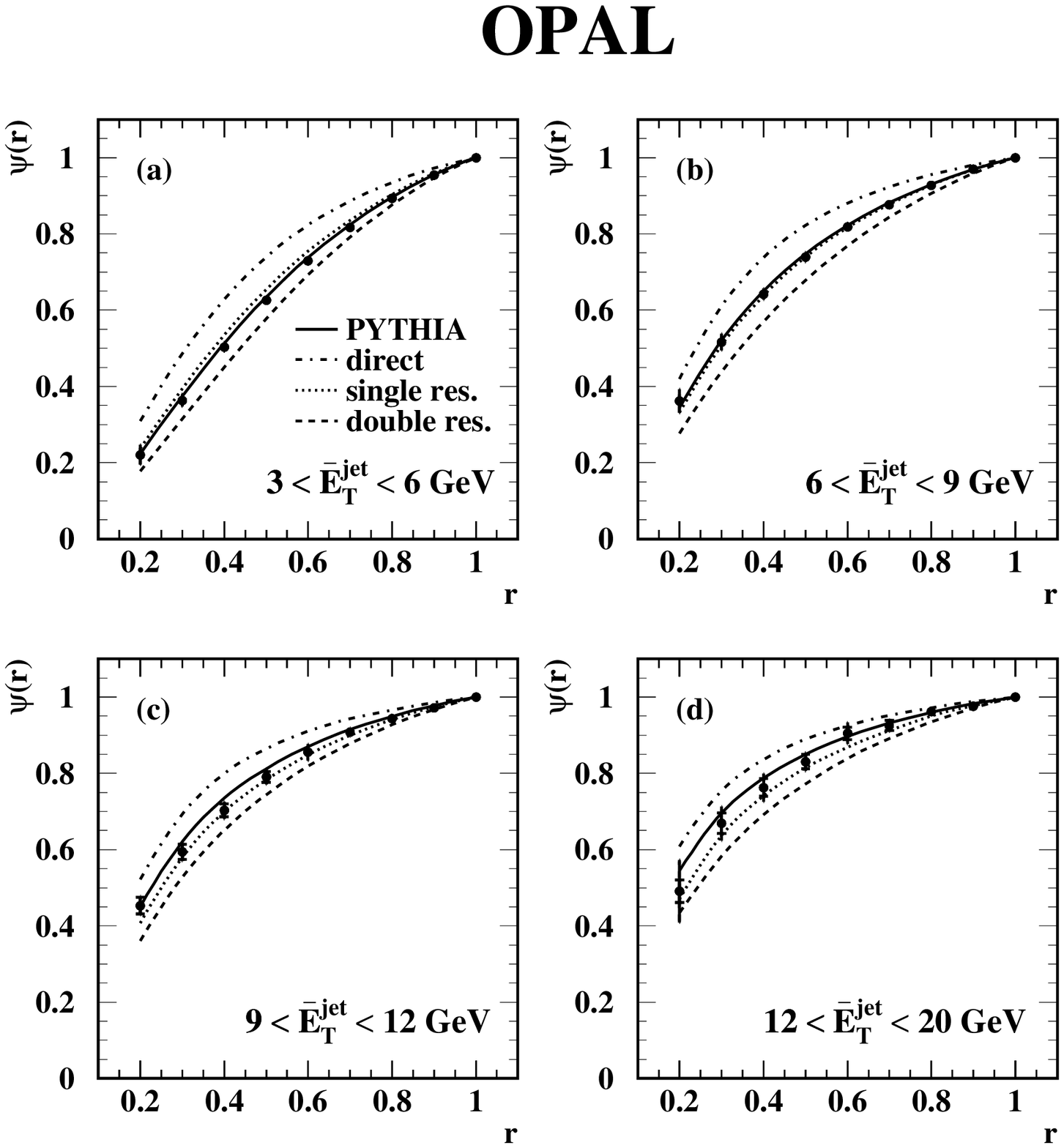}
           }
   \end{center}
\caption{The measured jet shapes, $\psi(r)$, corrected to the hadron 
 level for each of the two highest $\ETJET$ jets. 
 The jet shapes are shown in bins of $\ETBAR$;
 (a) $3<\ETBAR<6$ GeV, (b) $6<\ETBAR<9$ GeV, (c) $9<\ETBAR<12$ GeV and
 (d) $12<\ETBAR<20$ GeV.The predictions of the direct, single-resolved and
 double-resolved processes and their sum 
 as predicted by PYTHIA are shown.
 The inner error bar shows the statistical error
and the outer error bar the statistical and systematic errors added
in quadrature.}
\label{figet2}
\end{figure}

The $\ETBAR$ dependence of the jet shapes in di-jet production
is presented in Figure~\ref{figet2}. 
The predictions of PYTHIA for direct, single-resolved and
double-resolved processes and their sum are compared to the measured
jet shapes. 
The central value of the plotted data points is obtained
with the correction calculated from PYTHIA.
The systematic error was obtained from the variation of the ECAL energy
by $\pm  5 \%$ and
from the difference of the results
from PYTHIA and PHOJET. 
The relative contribution of the processes in PYTHIA has also been changed
over a range consistent with what the match to Figure~\ref{fig-xg}
will allow, and the resulting changes in the corrections have been added
to the systematic error.
The jets become
narrower as $\ETBAR$ increases. The predictions of PYTHIA with
the default relative contributions of the different processes using the
SaS-1D parametrisation reproduce the data reasonably well.
There is almost no difference between the predictions of PYTHIA and
PHOJET using GRV (not shown). 
The differences of the jet shapes for direct, single- and double-resolved
processes are expected to be due to the different fractions of quark and
gluon jets. It has been shown that gluon jets are broader than
quark jets~\cite{bib-coneopal}.
According to the prediction of PYTHIA using SaS-1D the fraction
of gluon jets is $12 \%$ for direct events, $16 \%$ for single-resolved events
and $41 \%$ for double-resolved events.

\begin{figure}[htbp]
   \begin{center}
      \mbox{
          \epsfxsize=16.7cm
          \epsffile{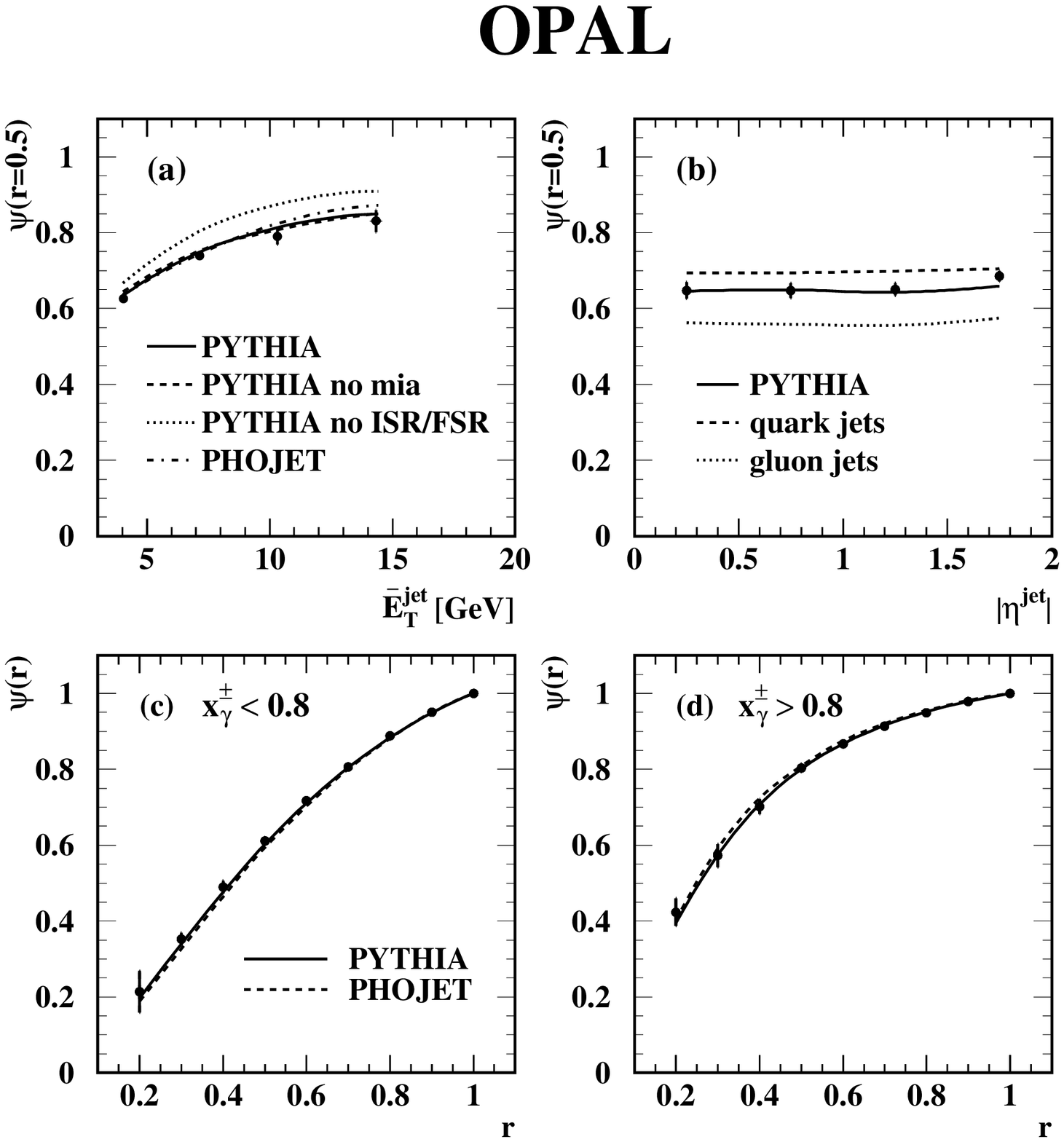}
           }
   \end{center}
\caption{The fraction of the transverse energy of the jets inside a cone of
radius $r=0.5$ 
around the jet axis is shown (a) as a function of $\ETBAR$ and (b) as
a function of $\etajet$. The measured jet shapes corrected to the hadron 
level, $\psi(r)$, are shown in (c) for
$\xgpm < 0.8$ and in (d) for $\xgpm > 0.8$. 
The statistical error is smaller than the symbol size. The error bars show
the statistical and systematic errors added in quadrature.}
\label{figshp2}
\end{figure}

The fraction of the transverse energy of the jets inside a cone of
radius $r=0.5$ around the jet axis, $\psi(r=0.5)$ is shown
as a function of $\ETBAR$
(Figure~\ref{figshp2}a) and as a function of
$|\etajet|$ (Figure~\ref{figshp2}b).
The position of the data points is the mean value of $\ETBAR$ in each
bin.
In Figure~\ref{figshp2}a the data are compared to PYTHIA with and without
multiple interactions and to PYTHIA without initial (ISR) and final state
QCD radiation (FSR). The PHOJET prediction is also shown.
The difference between PYTHIA with and without multiple interactions (mi)
is very small, whereas the PYTHIA prediction without initial and final state
QCD radiation, where only fragmentation effects have been taken into account,
leads to jets which are significantly narrower. This difference increases with
increasing $\ETBAR$.
Fig.~\ref{figshp2}b shows that there is no observed 
dependence of the jet shape
on $|\etajet|$. The prediction of PYTHIA is in good agreement with the data.
Gluon jets are predicted to be broader than quark jets.
The hadron jets in the Monte Carlo events have been identified as a quark
or gluon jet depending on the type (quark or gluon) of the closest parton
in the $\eta\phi$ plane.

The results for $\psi(r)$ are presented in Figure~\ref{figshp2}c and d for
both
$x^{\pm}_{\gamma}$ smaller and larger than 0.8. It is observed that the 
measured jet shapes for $x^{\pm}_{\gamma} < 0.8$, where more gluon jets are
expected, are broader than those for 
$x^{\pm}_{\gamma} > 0.8$. For both regions of $x^{\pm}_{\gamma}$ the
Monte Carlo generators reproduce the data reasonably well.

\section{Inclusive two-jet cross-sections and NLO calculations}
\label{sec-cross}
To obtain absolute jet cross-sections which can be
compared to theoretical calculations, we use the Monte Carlo simulation and
an unfolding program~\cite{bib-blobel} to correct
for the selection cuts, the resolution effects
of the detector and the background from
non-signal processes.
The same technique was used as described in Ref.~\cite{bib-opalgg}.
To improve the performance of the unfolding program in
the region $\ETJET>3$~GeV, bin-to-bin migration effects from jets at
low $\ETJET$ must be taken into account.
Therefore the jets are actually found with $\ETMIN=2$~GeV
and the unfolding is performed in the full $\ETJET>2$~GeV range,
but the unfolded jet cross-section are only shown for $\ETJET>3$~GeV.
The uncorrected
number of jets reconstructed in the range $3<\ETJET<4$~GeV is about
$15\%$ larger with $\ETMIN=3$~GeV than with $\ETMIN=2$~GeV.
This difference decreases to less than $5\%$ for higher $\ETJET$.

\begin{figure}[htbp]
   \begin{center}
      \mbox{
          \epsfxsize=16.7cm
          \epsffile{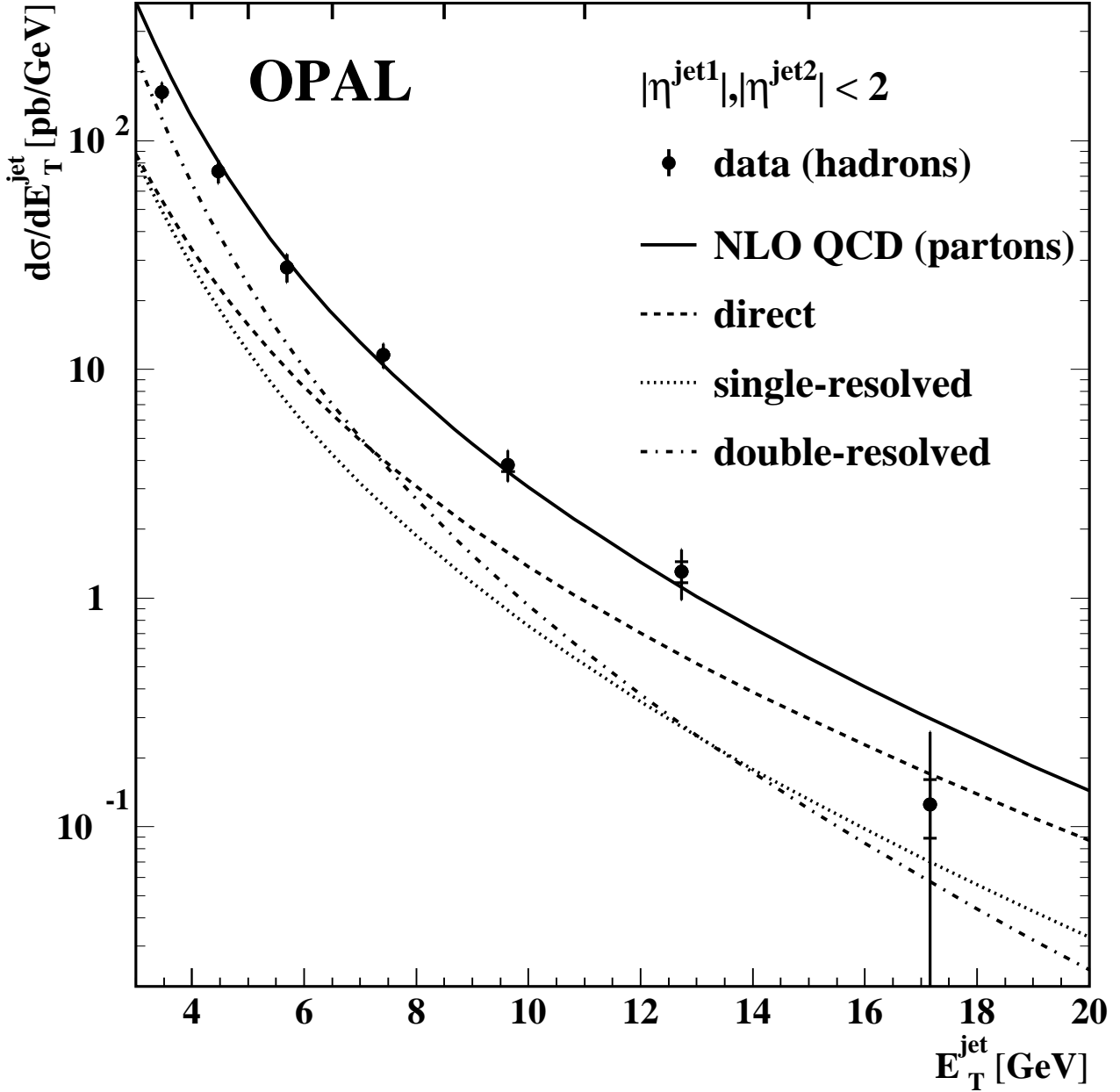}
           }
   \end{center}
\caption{The inclusive two-jet cross-section as a function
of $\ETJET$ for events with $|\etajet|<2$ compared to the NLO
calculation by Kleinwort and Kramer \protect\cite{bib-kleinwort}.
The direct, single-resolved and double-resolved cross-sections
and the sum (continuous line) are shown separately. The inner error bar
shows the statistical error and the outer error bar the statistical
and systematic errors added in quadrature.}
\label{fig-ettwojet}
\end{figure}

In Fig.~\ref{fig-ettwojet}, the inclusive two-jet cross-section is
shown as a function of $\ETJET$.
The average transverse energy, $\langle\ETJET\rangle$,
within each bin, which is plotted on the abscissa, is determined
as proposed in Ref. \cite{bib-stick}.
It is obtained
by integrating an exponential
function which is fitted to the neighbouring data points. 
The error on $\langle\ETJET\rangle$ is calculated by varying the
slope of the exponential function.
The  error bars show the statistical and the systematic errors,
calculated in the same way as in Section~\ref{sec-angdis}, and
an additional error from the unfolding procedure are
added in quadrature. 
The bin sizes, which are indicated by
the tic marks at the top of the Figures, approximately reflect the
experimental resolution.
The results are summarised in Table \ref{tab-twojet}.

The $\ETJET$ distribution is compared to a NLO
perturbative QCD calculation of the inclusive two-jet cross-section
by Kleinwort and Kramer \cite{bib-kleinwort} who use
the NLO GRV parametrisation of the parton distribution functions of the
photon~\cite{bib-grv}.
Their calculation was repeated for the kinematic conditions
of this analysis.
The renormalisation and factorisation 
scales are chosen to be equal to $\ETJET$. The scale dependence
of the NLO QCD calculations is expected to be small~\cite{bib-klasen2}.
The strong coupling
constant $\alpha_{\rm s}$ is calculated from the two-loop formula with 
$\Lambda^{(5)}_{\overline{\rm MS}}=130$ MeV, since
this value is also used in the NLO-GRV parametrisation.
Changing $\Lambda^{(5)}_{\overline{\rm MS}}$ from 130 to 250 MeV
only in the $\alpha_{\rm s}$ formula 
increases the two-jet cross-section by factors from 1.4
to 1.07 in the range $3<\ETJET<16$~GeV for 
$\sqee=130-136$~GeV~\cite{bib-lambda}.

The data points are in good agreement with the calculation except
in the first bin where the calculation predicts a much higher cross-section.
The symmetric cuts on $\ETJET$ lead to singularities of the
NLO calculations. This problem only affects the first bin, where
the NLO calculations are not reliable.
The NLO QCD calculation gives the jet cross-section
for massless partons, whereas the 
experimental jet cross-sections are measured for hadrons. 
The uncertainties due to the modelling of the hadronisation process
have not been taken into account.
Because the partons in the Monte Carlo models
and the partons in the NLO calculations are defined in different ways it
is impossible to use the Monte Carlo to correct the data so that it can be
compared with the NLO parton level predictions. 
If PYTHIA had been used to calculate a correction we would have had to
increase the cross-section by a factor of between 1.2 and 1.3, with the
largest effects at low $\ETJET$.

The predictions for the 
direct, single- and double-resolved parts and their sum are
shown separately. The resolved cross-sections is the largest component in
the region $\ETJET\;\sleq\;8$~GeV, whereas, at high $\ETJET$
the direct cross-section is largest.
\section{Influence of the underlying event}
The NLO QCD calculations also do not take into account the possibility
of an underlying event which leads to an increased
jet cross-section. The underlying event is simulated in the Monte Carlo models
PYTHIA and PHOJET which will be used to compare to
different LO parametrisations of
the parton distribution, SaS-1D~\cite{bib-sas}, GRV~\cite{bib-grv}
and LAC1~\cite{bib-LAC1}. These sets of parton distributions contain different
parametrisations of the gluon density with LAC1
predicting a much larger gluon density than GRV and SaS-1D.
In PYTHIA and PHOJET the modelling of the
underlying event includes multiple interactions.
The contribution from multiple interactions has to be tuned
using quantities which are not directly correlated to the jets,
since otherwise effects of the parton distributions and
of the underlying event cannot be distinguished.
A significant difference between the
predicted two-jet cross-sections obtained with PYTHIA and PHOJET
using the same parametrisation of the parton distributions
was observed in our studies at lower energies~\cite{bib-opalgg}.
By adjusting the cutoff parameter $\PTMIA$ for multiple interactions
the model dependence should decrease significantly. 

It is expected that the transverse energy flow outside the jets
measured as a function of $\xg$
is correlated to the underlying event~\cite{bib-h1}.
No effect due to the underlying event
is expected for direct events at large $\xg$.
The increase of the transverse energy flow outside
the two jets at small $\xg$ can therefore be used
to tune the number of multiple interactions in the model.

\begin{figure}[htbp]
   \begin{center}
      \mbox{
          \epsfxsize=16.7cm
          \epsffile{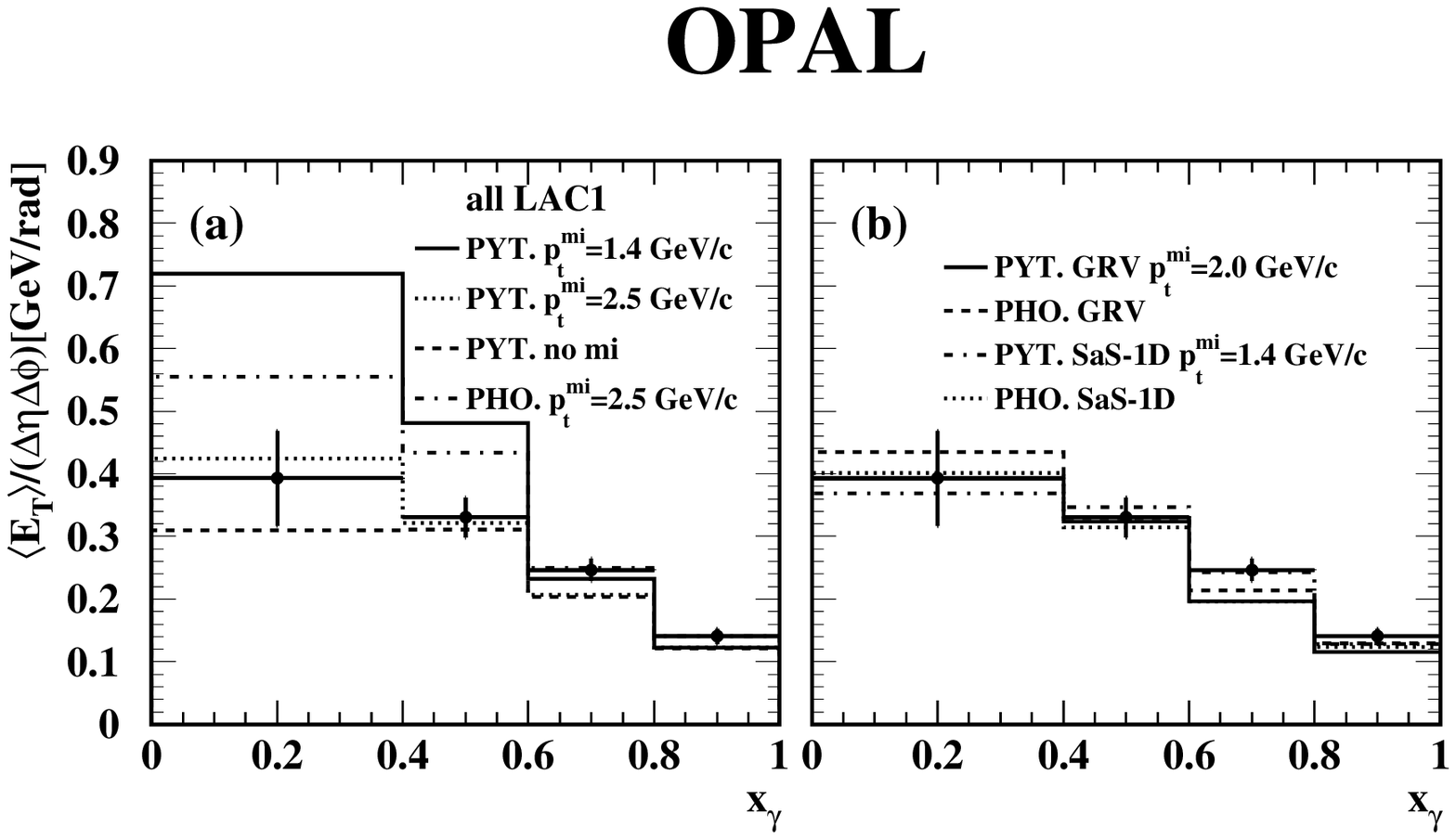}
           }
   \end{center}
\caption{(a) Transverse energy flow outside the jets in the central
 rapidity region $|\eta^*| < 1$ as a function of $\xg$. 
The statistical error is smaller
 than the symbol size. The error bars show the statistical
 and systematic errors added in quadrature.}
\label{fig-mia}
\end{figure}

The events were boosted into their
centre-of-mass system and the transverse energy flow was measured
as a function of $\xg$ in the central rapidity region $|\eta^*|<1$. 
The regions around the jet axes with $R<1.3$ are excluded from the
energy sum.
As in Section~\ref{sec-xg}, $\xg$ denotes that the transverse energy flow
of each event is added to the plot at the values of $\xgp$ and of $\xgm$.
Fig.~\ref{fig-mia} shows the transverse
energy flows corrected to the hadron level.
The systematic error was obtained from the difference of the results
from PYTHIA and PHOJET and from the variation of the ECAL energy
by $\pm  5 \%$.
Fig.~\ref{fig-mia}a shows the results of
PYTHIA using the LAC1 parametrisation with different
$\PTMIA$ cutoff parameters. 
The transverse energy flow for the default $\PTMIA$ of 1.4 GeV$/c$ 
is much too high in the first $\xg$ bins. 
Without multiple 
interactions the transverse energy flow is too low. 
An optimised value of $\PTMIA= 2.5$~GeV$/c$
leads to a reasonable description of the data.
The average number $n_{\rm mi}$ of interactions in double-resolved
di-jet events is 4.1 for LAC1 with $\PTMIA= 1.4$~GeV$/c$ and
1.5 with $\PTMIA= 2.5$~GeV$/c$.

The difference of the transverse energy flow using SaS-1D in PYTHIA
with and without multiple interactions  
($\PTMIA=1.4$~GeV$/c$) is
very small and the predicted transverse energy flow is in good agreement
with the data. The best description using PYTHIA
and GRV is obtained with a $\PTMIA$ of 2.0~GeV$/c$.
For all further comparisons with PYTHIA,
the cutoff parameter $\PTMIA$ was set to $2.5$~GeV$/c$
for LAC1, to $2.0$~GeV$/c$ for GRV and to $1.4$~GeV$/c$ for SaS-1D.

In $\gamma$p collisions at HERA the GRV parametrisation with
a cutoff parameter of $\PTMIA=1.2$~GeV$/c$ for PYTHIA 
has been found to be in good agreement
with the data, whereas the cutoff parameter was set to $2.0$~GeV$/c$
for LAC1~\cite{bib-h1}. With this choice the models slightly overestimate
the transverse energy flows at low $\xg$ in our data. 

PHOJET with either SaS-1D or GRV is in reasonable agreement with the data.
Changing the default cutoff of PHOJET from $\PTMIA= 2.5$~GeV$/c$ 
does not affect the transverse energy flow significantly,
the parameter $\PTMIA$ was therefore left unchanged in PHOJET.  

\section{Inclusive two-jet cross-sections as function of 
\boldmath $|\etajet|$ \unboldmath}
\begin{figure}[htbp]
   \begin{center}
      \mbox{
          \epsfxsize=16.7cm
          \epsffile{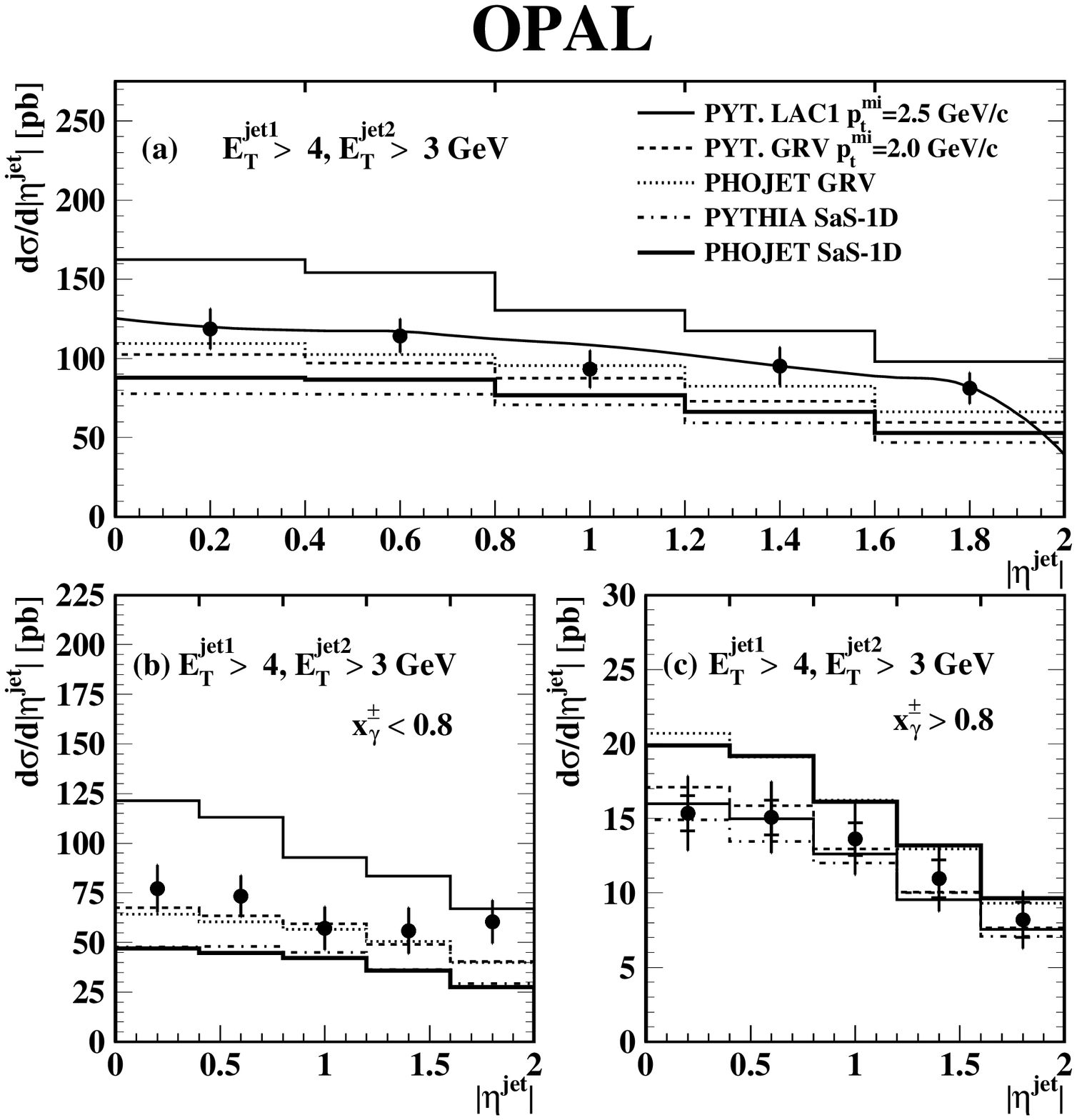}
           }
   \end{center}
\caption{The inclusive two-jet cross-section as a function of $|\etajet|$
 for events with $E^{\rm jet1}_{\rm T}> 4$~GeV and
 $E^{\rm jet2}_{\rm T}> 3$~GeV are
 shown for (a) all events and (b) for events with a large contribution of
 double-resolved events by requiring $\xgpm < 0.8$ and (c) for events
 with a large contribution of direct events by requiring $\xgpm > 0.8$.
 The curve in (a) shows the prediction of the NLO QCD calculation using
 the NLO GRV parametrisation. The inner error bar shows the statistical error
and the outer error bar the statistical and systematic errors added
in quadrature.}
\label{fig-etatwo}
\end{figure}

\begin{figure}[htbp]
   \begin{center}
      \mbox{
          \epsfxsize=16.7cm
          \epsffile{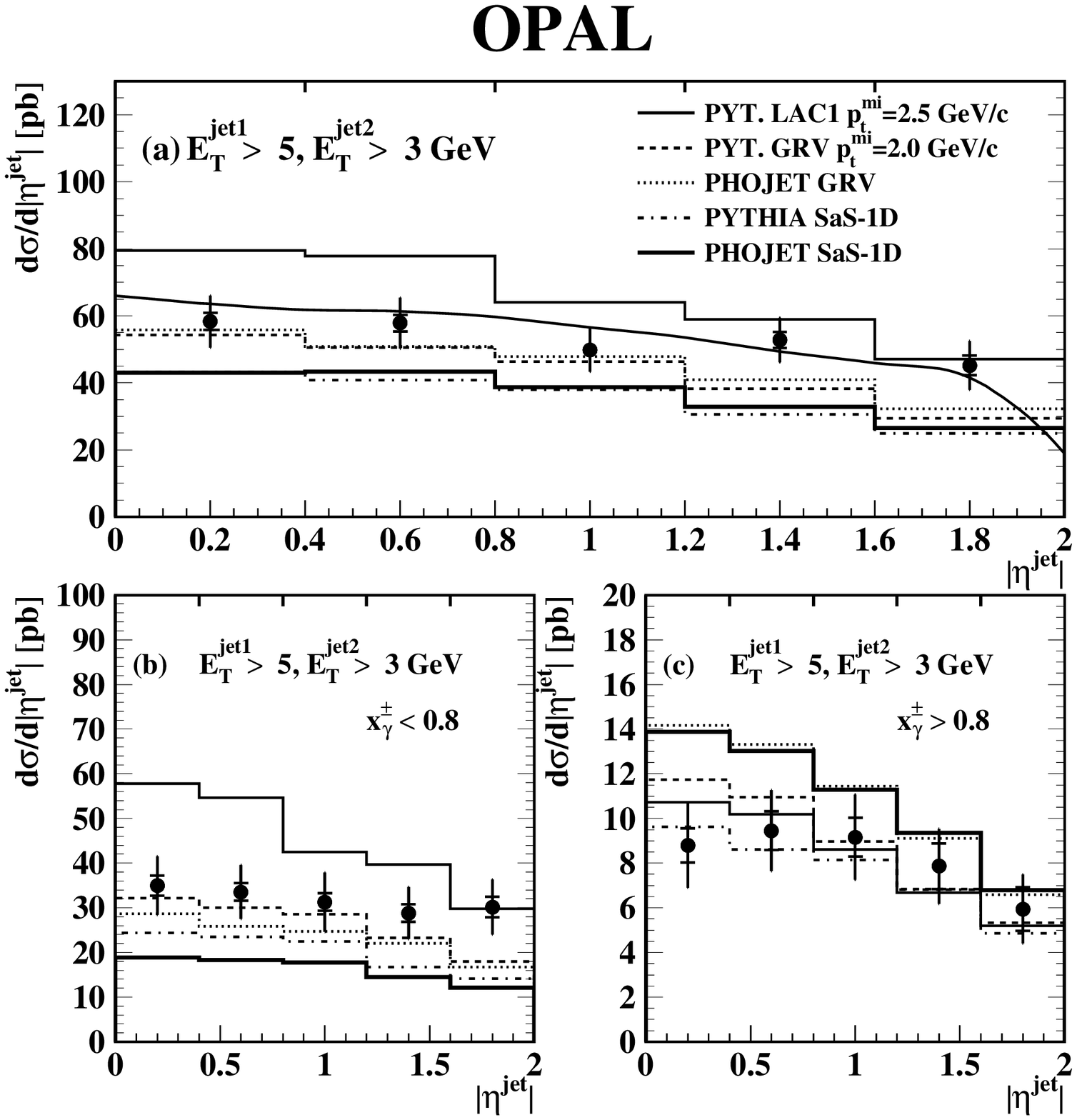}
           }
   \end{center}
\caption{The inclusive two-jet cross-section as a function of $|\etajet|$
 for events with $E^{\rm jet1}_{\rm T}> 5$~GeV and
 $E^{\rm jet2}_{\rm T}> 3$~GeV are
 shown (a) for all events and (b) for events with a large contribution of
 double-resolved events by requiring $\xgpm < 0.8$ and (c) for events
 with a large contribution of direct events by requiring $\xgpm > 0.8$.
 The curve in (a) shows the prediction of the NLO QCD calculation using
 the NLO GRV parametrisation. The inner error bar shows the statistical error
and the outer error bar the statistical and systematic errors added
in quadrature.}
\label{fig-etatwo5}
\end{figure}

The size and $|\etajet|$ dependence of the inclusive two-jet cross-section,
which is dominated by the low $\ETJET$ events, depend on the
chosen parton distribution functions 
which mainly differ in the assumptions on
the gluonic content of the photon. This leads to different predictions
for the inclusive two-jet cross-section, especially for
double-resolved events.
The inclusive two-jet cross-section as a function of $|\etajet|$
is shown in Figure~\ref{fig-etatwo} for events with
\linebreak
$E^{\rm jet1}_{\rm T}>4$~GeV and $E^{\rm jet2}_{\rm T}>3$~GeV and
in Figure~\ref{fig-etatwo5}
for events with $E^{\rm jet1}_{\rm T}>5$~GeV and
\linebreak $E^{\rm jet2}_{\rm T}>3$~GeV.
The asymmetric cuts have been applied because NLO calculations with
symmetric cuts are not infrared safe~\cite{bib-asym}.

The data sample is separated into events with a large contribution
from double-resolved processes by requiring $\xgpm<0.8$
(Fig.~\ref{fig-etatwo}b and \ref{fig-etatwo5}b) and into events
with a large contribution
from direct processes by requiring $\xgpm>0.8$
(Fig.~\ref{fig-etatwo}c and \ref{fig-etatwo5}c).
The results are summarised in
Tables~\ref{tab-etares} and~\ref{tab-etares5}.
The average $\langle|\etajet|\rangle$ values
are consistent with being at the centre of the bins.
Each jet is included with its value of $|\etajet|$ in
the cross-section measurement.
Within the statistical and systematic uncertainties
of the measurement, the data distributions are nearly independent
of $|\etajet|$ in Fig.~\ref{fig-etatwo}a and b
with a small drop towards $|\etajet|=2$, whereas the data distribution
of the direct events shown in Fig.~\ref{fig-etatwo}c at $|\etajet|=2$
falls to about half of its value  at $|\etajet|=0$.
In the kinematic range shown, this is in agreement with the expectations
of the Monte Carlo models.

The NLO QCD calculation of
the inclusive two-jet cross-section is in excellent agreement with
the differential cross-section ${{\mathrm d}\sigma}/{{\mathrm d}|\etajet|}$
shown in Figs.~\ref{fig-etatwo}a and~\ref{fig-etatwo5}a.
The SaS-1D parametrisation~\cite{bib-sas} with the PYTHIA and PHOJET
models predicts a two-jet cross-section
which is significantly too low for the whole data sample and for events
with a large contribution from double-resolved events ($\xgpm<0.8$).
The cross-section using GRV~\cite{bib-grv} is a bit too low for the whole data
sample but is able to describe the cross-section for the events
with $\xgpm<0.8$. In this range
the cross-sections as predicted using LAC1~\cite{bib-LAC1}.
are much too high. It should be noted that the overall normalisation
of jet cross-sections can alway be adjusted in a LO
calculation by changing the value of the strong coupling $\alpha_{\rm s}$.

In contrast to these discrepancies between the predicted
cross-sections for $\xgpm<0.8$ due to different parametrisations of the
parton distributions, the differences between the predicted
cross-sections for events with a large contribution of direct events 
($\xgpm>0.8$) seem to depend mainly on the model, PHOJET or PYTHIA,
and not the parametrisation used 
(Figs.~\ref{fig-etatwo}c and~\ref{fig-etatwo5}c). The PHOJET cross-section is
slightly too high, whereas PYTHIA describes the data well. As
expected, this is independent of the chosen cutoff parameter 
for multiple interactions.

\begin{figure}[htbp]
   \begin{center}
      \mbox{
          \epsfxsize=16.7cm
          \epsffile{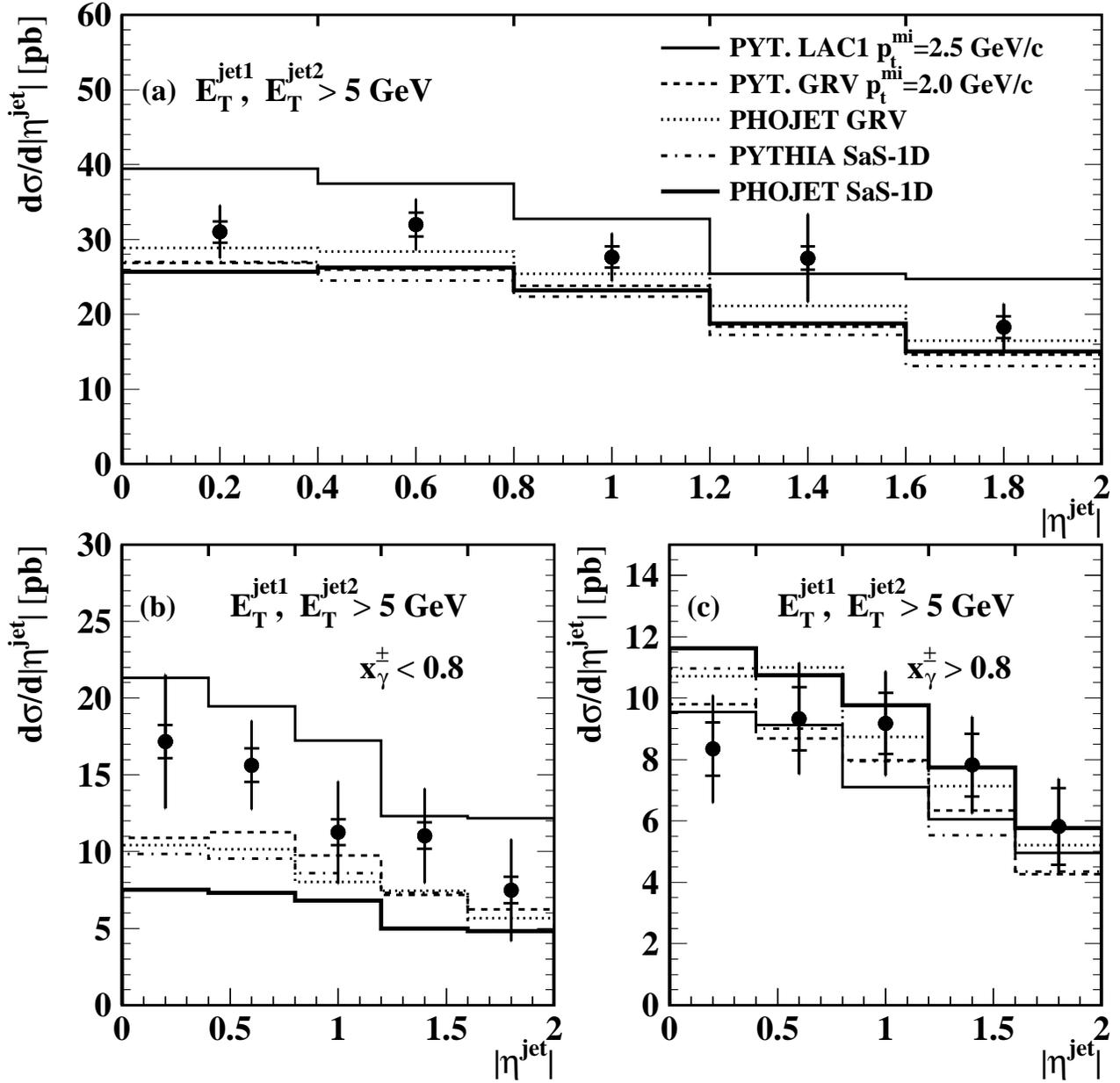}
           }
   \end{center}
\caption{The inclusive two-jet cross-section as a function of $|\etajet|$
 for events with $E^{\rm jet}_{\rm T}> 5$~GeV are
 shown (a) for all events and (b) for events with a large contribution of
 double-resolved events by requiring $\xgpm < 0.8$ and (c) for events with
 a large contribution of direct events by requiring $\xgpm > 0.8$.
 The inner error bar shows the statistical error and the outer error bar
 the statistical and systematic errors added in quadrature.}
\label{fig-etatwo55}
\end{figure}

To further
reduce the influence from multiple interactions and hadronisation effects
the inclusive two-jet cross-section was also measured for events 
with $\ETJET > 5$~GeV. 
(Fig.~\ref{fig-etatwo55}). The results are summarised in
Table~\ref{tab-etares55}. The fraction of resolved events is smaller
in this data sample and the difference between the different
parametrisations decreases. Nevertheless the LAC1 prediction
is still too high for the whole data sample and especially for the 
data sample with a large contribution of double-resolved events 
(Fig.~\ref{fig-etatwo55}b). 
The fraction of the cross-section with $\xgpm>0.8$ 
compared to the total jet cross-section is significantly 
larger in the range $E^{\rm jet}_{\rm T}> 5$~GeV than 
for the lower $E^{\rm jet}_{\rm T}$ ranges.

\section{Conclusions}
\label{sec-conclusions}
We have measured di-jet production in photon-photon interactions with
the OPAL detector at $\ee$ centre-of-mass energies $\sqee$ of 161
and 172~GeV with an integrated luminosity of 20~pb$^{-1}$.
Jets were identified using a
cone jet finding algorithm with $R=1$ in the kinematic range
$\ETJET>3$~GeV and $|\etajet|<2$.
Two-jet events originating mainly from direct and double-resolved
photon interactions were separated experimentally using
the variables $\xgp$ and $\xgm$.
The Monte Carlo models PYTHIA and PHOJET describe the 
transverse energy flow around the jets reasonably well.

The distribution of the parton scattering angle $\theta^*$
has been reconstructed from the rapidities
of the two jets. Data samples with large direct and double-resolved
contributions have been compared to LO and NLO QCD
calculations. 
A strong rise has been observed in the $|\cost|$ distribution
of the data sample with a large double-resolved contribution
at high $|\cost|$, as expected from QCD. The flatter $|\cost|$ distribution
of the data sample with a large contribution from direct events is also in
good agreement with the QCD calculation.

The energy profile of the jets has been measured in different regions
of $\ETBAR$. It has been observed that jets with high $\ETBAR$
are narrower than jets with small values of $\ETBAR$. The jet shape
in events with $\xgpm<0.8$, where double-resolved events are expected
to dominate, is found to be broader than the jet shape in events with
$\xgpm>0.8$, where direct events are expected to dominate.
These differences are assumed to be
caused by different fractions of quark jets and
gluon jets. The influence of multiple interactions and of different
parametrisations of the parton distribution functions of the photon
is small.

The inclusive two-jet cross-sections were measured as a function
of $\ETJET$ and $|\etajet|$. 
The measured cross-sections are in good agreement
with next-to-leading order QCD calculations by
Klasen, Kleinwort and Kramer~\cite{bib-kleinwort} above $\ETJET = 4$~GeV
using the NLO GRV parametrisation of the parton distributions of the photon.

The inclusive two-jet cross-section is dominated by the resolved
processes in the low $\ETJET$ region. 
In order to distinguish between the 
contributions to the jet cross-section from possible multiple interactions
between the spectator partons and from the parton densities, the contribution 
of multiple interactions in the models has been tuned
using the measured transverse energy flow outside the jets at low $\xg$.
However, within the errors of the
measurement we are unable to differentiate between models
with and without multiple interactions.

Using PYTHIA and PHOJET the LO GRV parametrisation is also able
to describe the two-jet cross-section
whereas the cross-section predicted based on the
SaS-1D parametrisation is too low and the prediction based on the LAC1
parametrisation is significantly too high. The same behaviour is
observed using a data sample with a large contribution from
double-resolved events. As expected, there is no sensitivity
to the choice of parametrisation for the complementary data
sample with a large contribution from direct events.
This behaviour still holds if the inclusive two-jet cross-sections
are measured for $E^{\rm jet}_{\rm T}> 5$~GeV.

\vspace{-4mm}
\section*{Acknowledgements}
We thank R.~Engel and T.~Sj\"ostrand for providing the
Monte Carlo code and for many useful discussions
and M.~Klasen, T.~Kleinwort and G.~Kramer for providing the NLO
calculations.
We particularly wish to thank the SL Division for the efficient operation
of the LEP accelerator at all energies
and for their continuing close cooperation with
our experimental group.  We thank our colleagues from CEA, DAPNIA/SPP,
CE-Saclay for their efforts over the years on the time-of-flight and trigger
systems which we continue to use.  In addition to the support staff at our own
institutions we are pleased to acknowledge the  \\
Department of Energy, USA, \\
National Science Foundation, USA, \\
Particle Physics and Astronomy Research Council, UK, \\
Natural Sciences and Engineering Research Council, Canada, \\
Israel Science Foundation, administered by the Israel
Academy of Science and Humanities, \\
Minerva Gesellschaft, \\
Benoziyo Center for High Energy Physics,\\
Japanese Ministry of Education, Science and Culture (the
Monbusho) and a grant under the Monbusho International
Science Research Program,\\
German Israeli Bi-national Science Foundation (GIF), \\
Bundesministerium f\"ur Bildung, Wissenschaft,
Forschung und Technologie, Germany, \\
National Research Council of Canada, \\
Research Corporation, USA,\\
Hungarian Foundation for Scientific Research, OTKA T-016660, 
T023793 and OTKA F-023259.\\

\newpage
\renewcommand{\arraystretch}{1.18}
\begin{table}[htbp]
  \begin{center}
    \begin{tabular}{|c|c|c|}
  \hline
$|\cost|$ & \multicolumn{2}{c|}{${\mathrm d}\sigma/{\mathrm d}|\cost|$~[pb]} \\ \cline{2-3}
    & resolved ($\xgpm < 0.8)$ & direct ($\xgpm > 0.8$)  \\ 
 \hline
  0.000 -- 0.106  &  $\pz6.2\pm1.7\pm0.3$ &  $\pz4.2\pm1.4\pm0.2$  \\
  0.106 -- 0.213  &  $\pz2.7\pm1.1\pm0.2$ &  $\pz5.0\pm1.5\pm0.4$  \\
  0.213 -- 0.319  &  $\pz9.7\pm2.1\pm1.1$ &  $\pz5.6\pm1.6\pm0.4$  \\
  0.319 -- 0.425  &  $10.3\pm2.2\pm2.3$   &  $\pz7.8\pm1.9\pm0.1$  \\
  0.425 -- 0.531  &  $10.5\pm2.2\pm1.7$   &  $\pz6.8\pm1.8\pm0.5$  \\
  0.531 -- 0.638  &  $22.7\pm3.3\pm2.4$   &  $\pz7.2\pm1.8\pm0.4$  \\
  0.638 -- 0.744  &  $30.1\pm3.8\pm2.5$   &  $\pz6.5\pm1.7\pm0.1$  \\
  0.744 -- 0.850  &  $51.9\pm5.0\pm3.6$   &  $15.8\pm2.7\pm0.7$    \\
  \hline
    \end{tabular}
    \caption{Differential two-jet cross-section as a function of $|\cost|$.
     The cross section is shown for events with $\xgpm<0.8$ and for events
     with $\xgpm>0.8$. The first error is statistical and the second error
     is systematic.}
    \label{tab-ang}
  \end{center}
\end{table}
\renewcommand{\arraystretch}{1.18}
\begin{table}[htbp]
  \begin{center} 
    \begin{tabular}{|c|c|r@{$\pm$}l@{$\pm$}r|}    
  \hline
 $\ETJET$  (GeV) &$\langle\ETJET\rangle$ (GeV) &  \multicolumn{3}{c|}{${\mathrm d}\sigma/{\mathrm d}\ETJET$~[pb/GeV]} \\
\hline 
 \pz3.0\pz -- \pz4.0 & $\pz3.47\pm0.01$ & 163\pzzz\pz & \pz3\pzz & 16\pzzz  \\ 
 \pz4.0\pz -- \pz5.0 & $\pz4.47\pm0.01$ &  73.6\pzz  & \pz2.0\pz  & 8.7\pz  \\ 
 \pz5.0\pz -- \pz6.5 & $\pz5.69\pm0.01$ &  27.9\pzz  & \pz0.9\pz  & 3.9\pz  \\
 \pz6.5\pz -- \pz8.5 & $\pz7.42\pm0.01$ &  11.5\pzz & \pz0.5\pz  & 1.4\pz  \\
 \pz8.5\pz -- 11.0   & $\pz9.64\pm0.01$ &   3.83 \pz & \pz0.26\pz & 0.60     \\
   11.0\pz -- 15.0   & $12.73\pm0.02$   &   1.30 \pz & \pz0.14\pz & 0.35     \\
   15.0\pz -- 20.0   & $17.16\pm0.02$   &   0.12 \pz & \pz0.04\pz & 0.13     \\
  \hline
   \end{tabular} 
    \caption{The inclusive two-jet cross-section as a function of $\ETJET$.
     The first error is statistical and the second error is systematic.}
    \label{tab-twojet}
  \end{center}
\end{table}
\renewcommand{\arraystretch}{1.18}
\begin{table}[htbp]
  \begin{center}
    \begin{tabular}{|c|c|c|c|}
  \hline
 $|\etajet|$ & \multicolumn{3}{c|}{${\mathrm d}\sigma/{\mathrm d}|\etajet|$~[pb]}    \\ \cline{2-4}
 & no $\xgpm$ cut & resolved ($\xgpm < 0.8$)  & direct ($\xgpm > 0.8$)  \\ 
 \hline
0.0 -- 0.4 & $118.7\pm3.7\pm12.8$ & $77.2\pm3.3\pm11.8$ & $15.3\pm1.2\pm2.5$ \\
0.4 -- 0.8 & $114.4\pm3.6\pm10.7$ & $73.3\pm3.1\pm10.4$ & $15.1\pm1.2\pm2.4$ \\
0.8 -- 1.2 & $\pz93.4\pm2.9\pm11.7$ & $57.2\pm2.5\pm10.9$&$13.6\pm1.1\pm2.4$ \\
1.2 -- 1.6 & $\pz95.1\pm3.5\pm11.7$ & $56.0\pm2.8\pm11.5$&$10.9\pm1.3\pm2.2$ \\
1.6 -- 2.0 & $\pz81.2\pm3.8\pm\pz9.7$ & $60.5\pm3.4\pm10.9$&$\pz8.2\pm1.2\pm1.9$ \\
  \hline
    \end{tabular}
    \caption{The inclusive two-jet cross-section as a function of $|\etajet|$
    for events with $E^{\rm jet1}_{\rm T}> 4$~GeV and
    $E^{\rm jet2}_{\rm T}> 3$~GeV.
    The inclusive two-jet cross-section is shown for all two-jet events and
    for data samples with a large contribution from double-resolved events
    by requiring $\xgpm < 0.8$ and with a large contribution from
    direct events by requiring $\xgpm > 0.8$. The first error is
    statistical and the second error is systematic.}
    \label{tab-etares}
  \end{center}
\end{table}
\renewcommand{\arraystretch}{1.18}
\begin{table}[htbp]
  \begin{center}
    \begin{tabular}{|c|c|c|c|}
  \hline
 $|\etajet|$ & \multicolumn{3}{c|}{${\mathrm d}\sigma/{\mathrm d}|\etajet|$~[pb]}    \\ \cline{2-4}
 & no $\xgpm$ cut & resolved ($\xgpm < 0.8$)  & direct ($\xgpm > 0.8$)  \\ 
 \hline
0.0 -- 0.4 & $58.3\pm2.6\pm7.7$ & $35.0\pm2.3\pm6.5$ & $8.8\pm0.8\pm1.9$ \\
0.4 -- 0.8 & $57.8\pm2.4\pm7.6$ & $33.6\pm2.0\pm6.0$ & $9.5\pm0.9\pm1.8$ \\
0.8 -- 1.2 & $49.8\pm2.0\pm6.5$ & $31.3\pm2.0\pm6.7$ & $9.2\pm0.9\pm1.9$ \\
1.2 -- 1.6 & $52.8\pm2.4\pm6.6$ & $28.8\pm2.0\pm5.8$ & $7.9\pm1.0\pm1.7$ \\
1.6 -- 2.0 & $45.2\pm2.9\pm7.2$ & $30.2\pm2.3\pm6.2$ & $5.9\pm1.0\pm1.5$ \\
  \hline
    \end{tabular}
    \caption{The inclusive two-jet cross-section as a function of $|\etajet|$
    for events with $E^{\rm jet1}_{\rm T}> 5$~GeV and
    $E^{\rm jet2}_{\rm T}> 3$~GeV.
    The inclusive two-jet cross-section is shown for all two-jet events and
    for data samples with a large contribution from double-resolved events
    by requiring $\xgpm < 0.8$ and with a large contribution from
    direct events by requiring $\xgpm > 0.8$. The first error is
    statistical and the second error is systematic.}
    \label{tab-etares5}
  \end{center}
\end{table}
\renewcommand{\arraystretch}{1.18}
\begin{table}[htbp]
  \begin{center}
    \begin{tabular}{|c|c|c|c|}
  \hline
 $|\etajet|$ & \multicolumn{3}{c|}{${\mathrm d}\sigma/{\mathrm d}|\etajet|$~[pb]}    \\ \cline{2-4}
 & no $\xgpm$ cut & resolved ($\xgpm < 0.8$)  & direct($\xgpm > 0.8$)  \\ 
 \hline
0.0 -- 0.4 & $31.0\pm1.4\pm3.5$ & $17.2\pm1.1\pm4.4$ & $8.3\pm0.9\pm1.7$ \\
0.4 -- 0.8 & $32.0\pm1.6\pm3.4$ & $15.6\pm1.1\pm2.9$ & $9.3\pm1.0\pm1.8$ \\
0.8 -- 1.2 & $27.6\pm1.4\pm3.2$ & $11.3\pm0.9\pm3.3$ & $9.2\pm1.0\pm1.7$ \\
1.2 -- 1.6 & $27.5\pm1.6\pm5.8$ & $11.0\pm0.9\pm3.1$ & $7.8\pm1.0\pm1.6$ \\
1.6 -- 2.0 & $18.3\pm1.5\pm3.1$ & $\pz7.5\pm0.9\pm3.3$ & $5.8\pm1.2\pm1.5$ \\
  \hline
    \end{tabular}
    \caption{The inclusive two-jet cross-section as a function of $|\etajet|$
    for events with $E^{\rm jet}_{\rm T}> 5$~GeV.
    The inclusive two-jet cross-section is shown for all two-jet events and
    for data samples with a large contribution from double-resolved events
    by requiring $\xgpm < 0.8$ and with a large contribution from
    direct events by requiring $\xgpm > 0.8$. The first error is
    statistical and the second error is systematic.}
    \label{tab-etares55}
  \end{center}
\end{table}

\newpage
\newpage
\newpage


\end{document}